\documentstyle[11pt,amssymb]{article}

\let\a=\alpha    
    
  \let\n=\nu

  \let\S=\Sigma   
\let\C=\Chi

\def\nn{\nonumber} \def\bd{\begin{document}} \def\ed{\end{document}}
\def\ds{\documentstyle} \let\fr=\frac \let\bl=\bigl \let\br=\bigr
\let\Br=\Bigr \let\Bl=\Bigl 
\let\bm=\bibitem
\let\na=\nabla
\let\pa=\partial \let\ov=\overline 
\newcommand{\be}{\begin{equation}} 
\newcommand{\ee}{\end{equation}} 
\def\ba{\begin{array}}
\def\ea{\end{array}}
\def\ft#1#2{{\textstyle{{\scriptstyle #1}\over {\scriptstyle #2}}}}
\def\fft#1#2{{#1 \over #2}}
\def\del{\partial}
\def\vp{\varphi}
\def\sst#1{{\scriptscriptstyle #1}}
\def\oneone{\rlap 1\mkern4mu{\rm l}}
\def\td{\tilde}
\def\wtd{\widetilde}
\def\ie{\rm i.e.\ }
\def\dalemb#1#2{{\vbox{\hrule height .#2pt
        \hbox{\vrule width.#2pt height#1pt \kern#1pt
                \vrule width.#2pt}
        \hrule height.#2pt}}}
\def\square{\mathord{\dalemb{6.8}{7}\hbox{\hskip1pt}}}
\newcommand{\ho}[1]{$\, ^{#1}$}
\newcommand{\hoch}[1]{$\, ^{#1}$}
\newcommand{\bea}{\begin{eqnarray}} 
\newcommand{\eea}{\end{eqnarray}} 
\newcommand{\ra}{\rightarrow}
\newcommand{\lra}{\longrightarrow}
\newcommand{\Lra}{\Leftrightarrow}
\newcommand{\ap}{\alpha^\prime}
\newcommand{\bp}{\tilde \beta^\prime}
\newcommand{\tr}{{\rm tr} }
\newcommand{\Tr}{{\rm Tr} } 
\def\0{{\sst{(0)}}}
\def\1{{\sst{(1)}}}
\def\2{{\sst{(2)}}}
\def\3{{\sst{(3)}}}
\def\4{{\sst{(4)}}}
\def\5{{\sst{(5)}}}
\def\6{{\sst{(6)}}}
\def\7{{\sst{(7)}}}
\def\8{{\sst{(8)}}}
\def\n{{\sst{(n)}}}
\def\cA{{{\cal A}}}
\def\cF{{{\cal F}}}
\def\tV{\widetilde V}
\def\tW{\widetilde W}
\def\tH{\widetilde H}
\def\tE{\widetilde E}
\def\tF{\widetilde F}
\def\tA{\widetilde A}
\def\im{{{\rm i}}}
\def\jm{{{\rm j}}}
\def\km{{{\rm k}}}
\def\tY{{{\wtd Y}}}
\def\ep{{\epsilon}}
\def\vep{{\varepsilon}}
\def\R{\rlap{\rm I}\mkern3mu{\rm R}}
\def\bD{{{\bar D}}}
\def\C{{{\Bbb C}}}
\def\H{{{\Bbb H}}}
\def\RP{{{\Bbb R}{\Bbb P}}} 
\def\CP{{{\Bbb C}{\Bbb P}}} 
\def\HP{{{\Bbb H}{\Bbb P}}}
\def\Z{{\Bbb Z}} 
\def\bfe{{\bf e}}
\def\bfq{{\bf q}}

\newcommand{\NP}{Nucl. Phys. }
\newcommand{\tamphys}{\it Center for Theoretical Physics,
Texas A\&M University, College Station, Texas 77843}
\newcommand{\ens}{\it Laboratoire de Physique Th\'eorique de l'\'Ecole
Normale Sup\'erieure\\
24 Rue Lhomond - 75231 Paris CEDEX 05}

\newcommand{\auth}{G.W. Gibbons\hoch{\sharp\,1,5}, 
H. L\"u\hoch{\ddagger\, 2}, 
C.N. Pope\hoch{\dagger\S\sharp\,1,3,4} and K.S. Stelle\hoch{\star\,4,5}}
\newcommand{\umich}{\it Department of Physics,
University of Michigan, Ann Arbor, Michigan 48109}

\begin{document}
\begin{flushright}
\hfill{DAMTP-2000-113,\ \ MCTP-01-38,\ \ 
CTP TAMU-27/01,\ \  Imperial/TP/00-01/29}\\

\hfill{\bf hep-th/0108191}\\
\hfill{August 2001}\\
\end{flushright}


\begin{center}
{\Large {\bf Supersymmetric Domain Walls from  
Metrics of Special Holonomy }}

\vspace{10pt}

\auth

\vspace{5pt}
{\hoch{\sharp} \it  DAMTP, Centre for Mathematical Sciences,
 Cambridge University, Wilberforce Road, Cambridge CB3 OWA, UK
}

\vspace{5pt}
{\hoch{\ddagger}\umich}

\vspace{5pt}
{\hoch{\dagger}\tamphys}

\vspace{5pt}
{\hoch{\S}\ens}

\vspace{5pt}
{\hoch{\star} \it The Blackett Laboratory, Imperial College\\
Prince Consort Road, London SW7 2BZ, UK}

\vspace{10pt}

\underline{ABSTRACT}
\end{center}

    Supersymmetric domain-wall spacetimes that lift to Ricci-flat
solutions of M-theory admit generalized Heisenberg (2-step nilpotent)
isometry groups. These metrics may be obtained from known
cohomogeneity one metrics of special holonomy by taking a ``Heisenberg
limit,'' based on an In\"on\"u-Wigner contraction of the isometry
group. Associated with each such metric is an Einstein metric with
negative cosmological constant on a solvable group manifold. We
discuss the relevance of our metrics to the resolution of
singularities in domain-wall spacetimes and some applications to
holography. The extremely simple forms of the explicit metrics suggest
that they will be useful for many other applications. We also give new
but incomplete inhomogeneous metrics of holonomy $SU(3)$, $G_2$ and
Spin(7), which are $T_1$, $T_2$ and $T_3$ bundles respectively over
hyper-K\"ahler four-manifolds.

{\vfill\leftline{}\vfill
\vskip  10pt
\footnoterule
{\footnotesize 
     \hoch{1} Research supported in part by the EC under RTN
contract HPRN-CT-2000-00122. \vskip -12pt} \vskip 14pt
{\footnotesize
        \hoch{2} Research supported in full by DOE Grant
DE-FG02-95ER40899. \vskip  -12pt} \vskip   14pt
{\footnotesize
        \hoch{3} Research supported in part by DOE 
grant DE-FG03-95ER40917. \vskip -12pt}  \vskip  14pt
{\footnotesize
        \hoch{4} Research supported in part by the EC under RTN
contract HPRN-CT-2000-00131. \vskip -12pt} \vskip 14pt
{\footnotesize
        \hoch{5} Research supported in part by PPARC SPG grant
PPA/G/S/1998/00613.  \vskip -12pt} \vskip 14pt

\pagebreak
\setcounter{page}{1}

\tableofcontents
\addtocontents{toc}{\protect\setcounter{tocdepth}{2}}
\newpage

\section{Introduction}

      Spaces with special holonomy are natural candidates for the
extra dimensions in string and M-theory, since they provide a simple
geometrical mechanism for reducing the number of
supersymmetries. Complete non-singular examples on non-compact
manifolds have been constructed where the Ricci-flat metrics can be
given explicitly.  Attention has mostly focused on cases of
cohomogeneity one that are asymptotically conical (AC) or asymptotically
locally conical (ALC).  The AC examples include the Eguchi-Hanson
metric in $D=4$ \cite{eguhan}, deformed or resolved conifolds in 
$D=6$ \cite{candossa}, and $G_2$ and Spin(7) holonomy metrics in $D=7$
and $8$ \cite{brysal,gibpagpop}.  Four-dimensional ALC solutions have
been also known for some time; they are the Taub-NUT \cite{hawk} and
Atiyah-Hitchin metrics \cite{ah}.  Supersymmetric higher-dimensional
ALC solutions have been elusive, until the recent explicit
constructions of ALC metrics with Spin(7) holonomy \cite{newspin7} and
$G_2$ holonomy \cite{bggg}.  One characteristic of those manifolds is
that they all have non-abelian isometry groups.

    Another situation where special holonomies are encountered is in
BPS solutions in lower-dimensional supergravities that are supported
by fields originating purely from the gravitational sector of a
higher-dimensional theory.  After oxidising the solutions back to the
higher-dimension, they give rise to Ricci-flat metrics.  Since the BPS
solutions partially break supersymmetry, while retaining a certain
number of Killing spinors, it follows that the Ricci-flat metrics will
have special holonomy.  In \cite{lav1,lav2,gibryc}, a geometrical
interpretation of these domain walls as Ricci-flat spaces with
toroidal fibre bundle level surfaces was given.

  Amongst the BPS solutions are a special class of domain-walls
($(D-2)$-branes in $D$ dimensions) that have the property of {\it
scale invariance}.  Technically, this means that they possess {\it
homotheties}, \ie conformal Killing symmetries where the conformal
scaling factor is a constant.

    The class of scale-invariant domain walls has appeared in another
context, namely the possibility of blowing up the singularities into
regular manifolds.  An example of this is given by a singular limit of
K3 that produces the transverse and internal dimensions of the
oxidation of an eight-dimensional 6-brane to $D=11$ \cite{gibryc}.
Since metrics on K3 are not known explicitly, the discussion was
necessarily a highly implicit one.  For our present purposes, however,
the salient properties of the K3 degeneration for this identification
with the domain wall were the appearance of a Heisenberg symmetry in
the singular limit, as well as a characteristic rate of growth of the
volume of the manifold as one recedes from the singularity.
Higher-dimensional examples with more internal directions, related to
higher-dimensional Calabi-Yau manifolds, were also considered in
\cite{gibryc}.  The associated domain walls have generalised
Heisenberg symmetries.  Since these Heisenberg groups are
homothetically invariant, they fall into the class of scale-invariant
domain walls that we are concerned with here.

   Four-dimensional supergravity domain walls arising from matter
superpotentials have been extensively studied in
\cite{cgr,cgs,cve89,cs1,cs2}.  Domain walls can also exist in maximal
supergravities.  For example, the D8-brane of massive type IIA
supergravity \cite{romans2a} was discussed in \cite{polwit}.
Generalised Scherk-Schwarz reductions give rise to lower-dimensional
massive supergravities that admit domain-wall solutions.  It was shown
that the D7-brane of type IIB and the D8-brane of massive type IIB are
T-dual, via a generalised Scherk-Schwarz $S^1$ reduction
\cite{berg78dual}.  A large class of domain walls arising from
Scherk-Schwarz reduction was obtained in \cite{clpstdomain}.  A
complete classification of such domain walls in maximal supergravities
was given in \cite{classp}.

    In lower-dimensional maximal supergravities, the ``cosmological
potential'' associated with the construction of supersymmetric domain
walls can arise either by generalised Kaluza-Klein reduction on
spheres, or by generalised toroidal reductions, where in both cases
internal fluxes are turned on.  The former give cosmological
potentials with at least two exponential terms, whilst the latter can
give potentials with a single (positive) exponential.  Importantly for
our purposes, the latter have the feature that the potential has no
intrinsic scale, and so the associated domain walls are scale
invariant.

   One motivation for the present work was to study the possibility of
smooth resolutions of Ho\v{r}ava-Witten type geometries. The idea
would be to seek everywhere smooth solutions of eleven-dimensional
supergravity that resemble two domain walls at the ends of a finite
interval. This was discussed in the context of domain walls based on
the $\frak{ur}$-Heisenberg group in Ref.\ \cite{gibryc}. In that
reference, it was shown that the singularity arising from the
vanishing of the harmonic function could be resolved by replacing the
four-dimensional hyper-K\"ahler metric ${\cal M}_4$ by a smooth
complete everywhere non-singular hyper-K\"ahler metric ${\cal M}^{\rm
resolved}$ on the complement in $\CP^2$ of a smooth cubic. The smooth
non-singular metric ${\cal M}^{\rm resolved}$ (called the BKTY metric
in \cite{gibryc}) is non-compact and has a single ``end'' (\ie a
single connected infinite region) which is given by ${\cal M}_4$ up to
small terms as one goes away from the domain-wall source.  This was
referred to in \cite{gibryc} as a ``single-sided domain wall.''

   The question naturally arises whether two such single-sided domain
walls may be joined together by an extended ``neck'' to form a
complete non-singular compact manifold ${\cal M}^{\rm compact}$ which
resembles the Ho\v{r}ava-Witten type geometry. For these purposes, we
need not restrict ourselves to four-dimensional manifolds and shall
consider any dimension less than eleven.
 
    To answer this question, we need to make some further assumptions
about the geometry of the neck region. In the light of the previous
example, it seems reasonable to require that the neck region be of
cohomogeneity one, perhaps with the group being one of the generalised
Heisenberg or Nilpotent groups that arise in the known supersymmetric
domain walls of M-theory. We could, of course, assume a more general
group or more generally drop the assumption that the neck region is
invariant under any group action. However, it does seem reasonable to
assume that the neck region is covered by a coordinate patch in which
the metric takes the form
\be
ds^2=dt^2+g_{ij}(x,t)dx^idx^j + \ldots \label{genmetric}
\ee
where $t$ is the proper distance along the neck.  In the cohomogeneity
one case $g_{ij}(x,t)dx^idx^j$ is a left-invariant metric on $G/H$,
and the ellipsis denotes extra terms which grow at very large $|t|$
and which may break $G$-invariance, corresponding to corrections to
the metric arising from the smooth resolutions at either end of the
interval.

     If ${\cal M}^{\rm compact}$ is Ricci-flat, or more generally if
$R_{tt}$ is non-negative, a simple consistency check immediately
arises.  The curves $x^i={\rm const}$, with tangent vectors
$T={\partial\over\partial t}$, constitute a congruence of geodesics of
the metric (\ref{genmetric}).  A congruence of curves in a
$(d+1)$-dimensional manifold is a $d$-dimensional family of curves,
one passing through every point of the manifold.  A congruence is {\it
hypersurface-orthogonal} (or vorticity-free) if the curves are
orthogonal to a family of $d$-dimensional surfaces.  The congruence we
are considering is clearly hypersurface-orthogonal, since every curve
is orthogonal to the surfaces $t=$constant.

 Now let
\be
V(x,t)=\sqrt{\det g_{ij}}
\ee
and let $\Theta={\dot V\over V}=g^{ij}{\partial g_{ij}\over \partial
t}$. Then $\Theta(t,x^i)$ is the expansion rate of the
geodesic congruence, and is therefore subject to the
Raychaudhuri equation,\footnote{For a brief review of the
Raychaudhuri equation, see Appendix 1.} which then reads
\be
{d\Theta\over dt}\le  -{1\over d}\, \Theta^2 -2\Sigma^2\,. 
\label{raychaudhuri}
\ee
where $\Sigma^2=\ft12\Sigma_{ij}\Sigma^{ij}$ and
$\Sigma_{ij}={\partial g_{ij}\over\partial t}-{1\over
d}g^{rs}{\partial g_{rs}\over\partial t}g_{ij}$ with $d+1={\rm dim}
{\cal M}^{\rm compact}$. The quantity $\Sigma^2$ is a measure of the
shear of the geodesic congruence given by $x^i={\rm const}$.  It is an
easy consequence of (\ref{raychaudhuri}) that if $\Theta$ is negative,
it remains negative and moreover tends to minus infinity in finite
proper time $t$.  This means that if the volume $V$ of the neck is
decreasing at one value of $t$,it is always decreasing. This simple
result, which may be verified in our explicit examples, indicates that
neck geometries in which $V$ increases as one goes outward to the
resolved regions in both directions are excluded. They also show that
resolving periodic arrays of domain walls in such a way as to make the
metric $g_{ij}(x,t)$ periodic in the proper distance variable $t$ are
excluded.

   In this paper, we shall study the relationship between
scale-invariant domain walls with Heisenberg symmetries, and complete
non-singular manifolds of special holonomy.  The simplest example,
which we discuss in section \ref{dwsec}, involves the oxidation of the 6-brane
in $D=8$ to $D=11$.  We show that the associated four-dimensional
Ricci-flat space (which has $SU(2)$ holonomy and thus is self-dual)
can be obtained from the Eguchi-Hanson metric, by taking a rescaling
limit in which the $SU(2)$ isometry degenerates to the Heisenberg
group.  (For this limit, one has to take the version of Eguchi-Hanson
where the curvature singularity appears in the manifold.  The relation
to the non-singular Eguchi-Hanson requires an additional analytic
continuation in the scale parameter.)  The other examples,
corresponding to higher-dimensional Ricci-flat metrics, are similarly
obtained as Heisenberg limits of higher-dimensional metrics of special
holonomy.

   It turns out that each of our scale-invariant Ricci flat metrics
with nilpotent isometry group acting on orbits of co-dimension one is
closely related to a complete homogeneous Einstein manifold with
negative cosmological constant with a solvable isometry group. The
simplest case is flat space, ${\Bbb E} ^n$ with metric $ds^2 = dt^2 +
dx^\mu\, dx_\mu$, which is related to the hyperbolic space $H^n$ with
metric $ds^2=dt^2 + e^{2k\, t}\, dx^\mu\, dx_\mu$.  In some cases
these metrics have been used as replacements for the hyperbolic space
$H^n$ in the AdS/CFT correspondence \cite{strom}. A striking feature
is the degenerate nature of the conformal boundary. For this reason we
shall include a discussion of these metrics and some of their
properties.
 
\section{Four-dimensional manifolds with $SU(2)$ holonomy}

\subsection{The basic domain-wall construction}\label{egsec}

    We consider a domain wall solution in eight-dimensional maximal
supergravity, supported by the 0-form field strength $\cF^1_{\0 23}$
coming from the dimensional reduction of the Kaluza-Klein vector in
$D=10$.\footnote{In this paper, we adopt the notation of
\cite{lpsol,cjlp1}, where the lower dimensional maximal supergravities
were obtained by consecutive $S^1$ reduction with the indices
$i=1,2,\ldots$ denoting the $i$'th coordinate in the reduction.}  The
metric is given by
\be
ds_8^2 = H^{1/6}\, dx^\mu\, dx_\mu + H^{7/6}\, dy^2\,,\label{d8dw}
\ee
where $H=1+m\, |y|$.  Oxidised back to $D=11$ using the standard KK
rules, we find that the eleven-dimensional metric is given by
$ds_{11}^2 = dx^\mu\, dx_\mu + ds_4^2$ where
\be
ds_4^2=  H\, dy^2 + H^{-1}\, (dz_1 + m\, z_3\, dz_2)^2 +
H\, (dz_2^2 + dz_3^2)\,.\label{4met}
\ee
Since the CJS field $F_\4$ is zero, $ds_4^2$ must be Ricci flat.  The
solution preserves $1/2$ of the supersymmetry, implying that
(\ref{4met}) has $SU(2)$ holonomy, \ie it is a Ricci-flat K\"ahler
metric.  The eleven-dimensional solution was obtained in \cite{lav2},
where domain wall charge quantisation through topological constraints
was discussed.  It was used in \cite{hullmassive} to argue that
M-theory compactified on a $T^2$ bundle over $S^1$ is dual to the
massive type IIA string theory.  The solution has a singularity at
$y=0$.  In \cite{gibryc}, it was shown that the metric (\ref{4met}) is
the asymptotic form of a complete non-singular hyper-K\"ahler metric
on the complement in $\CP^2$ of a smooth cubic curve.  The metric
(\ref{4met}) was obtained by means of a double T-duality
transformation of the D8-brane solution of the massive IIA theory in
\cite{berg78dual}.

   In the orthonormal basis
\bea
&&e^0 = H^{1/2}\,dy\,,\qquad
e^1 = H^{-1/2}\, (dz_1 + m\, z_3\, dz_2)\,,\nn\\
&&e^2 = H^{1/2}\, dz_2\,,\qquad e^3= H^{1/2}\, dz_3\,,\label{k3basis}
\eea
it is easily verified that the 2-form,
\be
J \equiv e^0\wedge e^1 -e^2\wedge e^3 = dy\wedge (dz_1 + m\, z_3\,
dz_2) - H\, dz_2\wedge dz_3\,,
\ee
is closed, and in fact covariantly constant.  It is a 
privileged K\"ahler form amongst the 2-sphere of complex structures.  

   If we define the holomorphic and antiholomorphic projectors $
P_i{}^j = \ft12(\delta_i^j -\im\, J_i{}^j)$ and 
$Q_i{}^j = \ft12(\delta_i^j +\im\, J_i{}^j)$, 
then complex coordinates $\zeta^\mu$ must satisfy the differential
equations
\be
Q_i{}^j\, \del_j\zeta^\mu=0\,.\label{complexcoords}
\ee
The integrability of these equations is assured from the fact that our
metric is already established to be K\"ahler.

    Let $x=y+\ft12 m\, y^2$, so that $H\, dy=dx$.  Then
(\ref{complexcoords}) can be shown to reduce to just the following
pair of independent equations:
\bea
\fft{\del\zeta^\mu}{\del x} + \im\, \fft{\del\zeta^\mu}{\del z_1}
&=&0\,,\nn\\
\fft{\del\zeta^\mu}{\del z_3} + \im\, \fft{\del\zeta^\mu}{\del z_2}
+\ft12 m\, z_3\, \Big(\fft{\del\zeta^\mu}{\del x} -
                   \im\, \fft{\del\zeta^\mu}{\del z_1}\Big) &=&0\,.
\eea
Solutions of these differential equations define the complex
coordinates $\zeta^\mu$ in terms of the real coordinates.

   Any pair of independent solutions to the above equations gives a
valid choice of complex coordinates.  A convenient choice is
\bea
\zeta_1 &=& z_3+\im\, z_2\,,\nn\\
\zeta_2 &=& x+\im\, z_1 -\ft14 m(z_2^2+z_3^2) +\ft{\im}{2}\, m \,
z_2\, z_3\,,
\eea
implying that the metric becomes
\be
ds_4^2 = H\, |d\zeta_1|^2 + H^{-1}\, |d\zeta_2 + \ft12 m\,
\bar\zeta_1\, d\zeta_1|^2\,,\label{4kahler}
\ee
with
\be
H= \Big[1+ m\, (\zeta_2+\bar\zeta_2) + \ft12 m^2\,
|\zeta_1|^2\Big]^{1/2}\,.
\ee
This agrees, up to coordinate redefinitions, with results in
\cite{gibryc}.  

   The metric in (\ref{4kahler}) has the characteristic Hermitean form
\be
ds^2 = 2g_{\mu\bar\nu}\, d\zeta^\mu\, d\bar\zeta^{\bar\nu}\,,
\ee
and in fact 
\be
g_{\mu\bar\nu} = \fft{\del^2 K}{\del\zeta^\mu\,
\del\bar\zeta^{\bar\nu}}\,,
\ee
where the K\"ahler function $K$ given by $K=2H^3/(3m^2)$.

    The metric (\ref{d8dw}) with $H=1+m\, |y|$ physically represents a
domain wall located at $y=0$.  This is constructed by patching two
sides, each of which is part of a smooth but incomplete metric in
which $H$ can instead be taken to have the form $H=m\, y$.  The metric
(\ref{4met}) with $H=m\, y$ has a scaling symmetry generated by the
dilatation operator
\be
D= y \, \fft{\del}{\del y} + 2 z_1\, \fft{\del }{\del z_1} 
+ z_2\, \fft{\del }{\del z_2} 
+ z_3\, \fft{\del }{\del z_3} 
\ee
As we shall discuss in section \ref{homothetysec}, this is a homothetic Killing
vector.  In addition, (\ref{4met}) is invariant under the linear
action of $U(1)$ on $(z_2,z_3)$, preserving the 2-form $dz_3\wedge
dz_2$.

\subsection{Domain-wall as Heisenberg contraction of 
Eguchi-Hanson}\label{egheissec}

   As discussed in \cite{gibryc}, the isometry group of the metric
(\ref{4met}) is the Heisenberg group, and it acts tri-holomorphically.
In other words, it leaves invariant all three of the 2-sphere's worth
of complex structures.  The Heisenberg group may be obtained as the
In\"on\"u-Wigner contraction of $SU(2)$.  It is not unreasonable,
therefore, to expect to obtain (\ref{4met}) as a limit of the
Eguchi-Hanson metric, which is the only complete and non-singular
hyper-K\"ahler 4-metric admitting a tri-holomorphic $SU(2)$ action.
One could consider the larger class of triaxial metrics admitting a
tri-holomorphic $SU(2)$ action considered in \cite{begipapo}, but our
metric is symmetric under the interchange of $z_2$ and $z_3$, and so
it is only necessary to consider the biaxial case.

    The Eguchi-Hanson metric is 
\be
ds_4^2 = \Big(1+\fft{Q}{r^4}\Big)^{-1}\, dr^2 + 
  \ft14 r^2\, \Big(1+\fft{Q}{r^4}\Big)\, \sigma_3^2 + \ft14 r^2\,
(\sigma_1^2 + \sigma_2^2)\,,\label{egha}
\ee
where the $\sigma_i$ are the left-invariant 1-forms
of $SU(2)$, satisfying 
\be
d\sigma_1 = -\sigma_2\wedge \sigma_3\,,\quad
d\sigma_2 = -\sigma_3\wedge \sigma_1\,,\quad
d\sigma_3 = -\sigma_1\wedge \sigma_2\,.\label{su2}
\ee
If we define rescaled 1-forms $\td \sigma_i$, according to
\be
\sigma_1=\lambda\, \td\sigma_1\,,\quad 
\sigma_2=\lambda\, \td\sigma_2\,,\quad 
\sigma_3=\lambda^2\, \td\sigma_3\,,\label{rescale}
\ee
then after taking the limit $\lambda\longrightarrow 0$,  we find that
(\ref{su2}) becomes\footnote{Heisenberg limits for more general
groups are discussed in Appendix B.} 
\be
d\td\sigma_1=0\,,\quad d\td\sigma_2=0\,,\quad d\td\sigma_3 =
-\td\sigma_1\wedge \td\sigma_2\,.\label{cartanm}
\ee
This is the same exterior algebra as in the 1-forms appearing in the
domain-wall metric (\ref{4met}), as can be seen by making the
associations
\be
\td\sigma_1 = m\, dz_2\,,\quad \td\sigma_2 = m\, dz_3\,,\quad
\td\sigma_3 = m\, (dz_1 + m\, z_3\, dz_2)\,.\label{sigz}
\ee

   To see how the Eguchi-Hanson metric (\ref{egha}) limits to the
domain-wall solution,  we should combine the rescaling (\ref{rescale})
with 
\be
r=\lambda^{-1}\, \td r\,,\qquad Q=\lambda^{-6}\, \wtd Q\,,
\ee
under which (\ref{egha}) becomes
\be
ds_4^2 = \Big(\lambda^4+\fft{\wtd Q}{\td r^4}\Big)^{-1}\, d\td r^2 + 
  \ft14 \td r^2\, \Big(\lambda^4+\fft{\wtd Q}{\td r^4}\Big)\, 
\td\sigma_3^2 + \ft14 \td r^2\,
(\td \sigma_1^2 + \td\sigma_2^2)\,.\label{egha2}
\ee
If we now define $\wtd Q = 16 m^{-4}$, $H=\ft14 m^2\, \td r^2$ and
take the Heisenberg limit $\lambda\longrightarrow 0$, we find, after
making the association (\ref{sigz}), that (\ref{egha2}) becomes
precisely (\ref{4met}) after a further coordinate change $y=\ft14
m\tilde r^2$.  

   The metric (\ref{egha}) has a curvature singularity at $r=0$. If
$Q<0$, this is not part of the Eguchi-Hanson manifold, which includes
only $r\ge -Q$. Intuitively, one may consider that the singularity is
hidden behind a bolt in this case. If $Q>0$, the singularity at $r=0$
is ``naked.'' In order to take the limit $\lambda\longrightarrow 0$ in the
rescaled metric (\ref{egha2}), we must let $\tilde Q>0$. The
near-singularity behaviour of the resulting metric is similar to that
of (\ref{egha}), but with the $SU(2)$ orbits flattened to Heisenberg
orbits.

\subsection{Heisenberg limit of the superpotential}\label{superpotentialsec}

   One may take the Heisenberg limit at the level of the equations of 
motion.   Thus the 4-dimensional metric
\be
ds_4^2 = (a\, b\, c)^2\, d\eta^2 + a^2\, \sigma_1^2 + b^2\, \sigma_2^2
+ c^2\, \sigma_3^2\,,
\ee
where
$a$, $b$ and $c$ are functions of $\eta$, will be Ricci-flat if
$\a\equiv\log a$, $\beta \equiv \log b$ and $\gamma\equiv \log c$ 
satisfy the equations of motion coming from the Lagrangian
\be
L = \dot\a\, \dot\beta + \dot\beta\, \dot\gamma + \dot\gamma\, \dot\a
-V\,,
\ee
with
\be
V = \ft14( a^4+b^4+c^4 - 2 b^2\, c^2 - 2 c^2\, a^2 -2 a^2\, b^2)\,.
\ee
A superpotential is given by
\be
W= a^2+b^2+c^2  - 2\lambda_1\, b\, c
- 2 \lambda_2\, c\, a - 2\lambda_3\, a\, b\,,
\ee
for any choice of the constants $\lambda_i$ that satisfy the three
equations
\be
\lambda_1=\lambda_2\, \lambda_3\,,\qquad \lambda_2= \lambda_3\,
\lambda_1\,,\qquad \lambda_3=\lambda_1\, \lambda_2\,.
\ee

   If $\lambda_1=\lambda_2=\lambda_3=0$, we get the equations of
motion for hyper-K\"ahler metrics with tri-holomorphic $SU(2)$ action,
solved in \cite{begipapo}.  It is possible to rescale the variables
$a$, $b$ and $c$ to obtain a superpotential giving the equations of
motion for metrics admitting a tri-holomorphic action of the
Heisenberg group.  One sets
\be
a= \lambda^{-1}\, \td a\,,\quad 
b= \lambda^{-1}\, \td b\,,\quad 
c= \lambda^{-2}\, \td c\,,
\ee
together with $\eta = \lambda^4\, \td\eta$ and $W=\lambda^{-4}\, \wtd
W$, giving
\be
\wtd W = \td c^2\,.
\ee
This gives rise to the first-order equations
\be
\fft{d\td\a}{d\td\eta} = \fft{d\td\beta}{d\td\eta}
=-\fft{d\td\gamma}{d\td\eta} = e^{2\td\gamma}\,,
\ee
from which one can easily rederive the domain-wall solution
(\ref{4met}).  

\subsection{Hypersurface-orthogonal homotheties}\label{homothetysec}

   The large-radius behaviour of the Eguchi-Hanson metric is that of a
Ricci-flat cone over $\RP^3$ \cite{begipapo}.  In this limit, the
metric becomes scale-invariant: scaling the metric by a constant
factor $\lambda^2$ is equivalent to performing the diffeomorphism
$r\longrightarrow \lambda\, r$.  This transformation is generated by
the ``Euler vector''
\be
E\equiv r\, \fft{\del}{\del r}\,,\label{eulerv}
\ee
which satisfies
\be
\nabla_\mu\, E_\nu = g_{\mu\nu}\,.
\ee
The vector $E^\mu$ is a special kind of conformal Killing vector $K^\mu$,
which would in general satisfy
\be
\nabla_\mu\, K_\nu + \nabla_\nu\, K_\nu = 2f\, g_{\mu\nu}\,.
\ee
If $f$ is constant, then $K^\mu$ generates a homothety, and if
$\nabla_\mu\, K_\nu=\nabla_\nu\, K_\mu$ then $K_\mu$ is a gradient,
and hence it is hypersurface-orthogonal.  The Euler vector $E^\mu$ in
(\ref{eulerv}) is an example of such a hypersurface-orthogonal
homothetic conformal Killing vector \cite{gibryc2}.

    The Heisenberg limit of the Eguchi-Hanson metric is also
scale-invariant.  In other words, the metric near the singularity is
scale-free.   Thus it is unchanged, up to a diffeomorphism, by
the transformation 
\be
z_1\longrightarrow \lambda^2\, z_1\,,\quad
z_2\longrightarrow \lambda\, z_2\,,\quad
z_3\longrightarrow \lambda\, z_3\,,\quad
y\longrightarrow \lambda\, y\,,\label{homo2}
\ee
where in this section we are taking $H=m\, y$.  This is generated by
the homothetic Killing vector
\be
D=y\, \fft{\del}{\del y} + 2z_1\, \fft{\del}{\del z_1} + 
z_2\, \fft{\del}{\del z_2} + z_3\, \fft{\del}{\del z_3}
\,.\label{eulerv2}
\ee
In contrast to the Euler vector (\ref{eulerv}) for the cone, the
homothety (\ref{eulerv2}) is neither a gradient nor 
is it proportional to a gradient, and so it is not
hypersurface-orthogonal.  

   The existence of the homothety generated by (\ref{homo2}) depends
crucially on the fact that the Heisenberg algebra, represented in
the Cartan-Maurer form in (\ref{cartanm}), is invariant under 
the scaling 
\be
\td\sigma_1\longrightarrow \lambda\, \td\sigma_1\,,\quad
\td\sigma_2\longrightarrow \lambda\, \td\sigma_2\,,\quad
\td\sigma_3\longrightarrow \lambda^2\, \td\sigma_3\,.\label{hscal}
\ee
By contrast, the original $SU(2)$ algebra, represented in (\ref{su2}),
is of course not invariant under (\ref{hscal}).   

   In \cite{gibryc2}, it was shown that (\ref{4met}) arises as the
large-distance limit of a non-compact Calabi-Yau 3-fold.  In the
large-distance limit, the metric becomes scale-invariant.

\subsection{Multi-instanton construction}\label{stacksec}

   The four-dimensional metric (\ref{4met}) can be related to the
general class of multi-instantons obtained in \cite{gibhaw}, which are
constructed as follows.  Let $x^i$ denote Cartesian coordinates on
$\R^3$.  We the write the metric
\be
ds_4^2 = V^{-1}\, (d\tau + A_i\, dx^i)^2 + 
V\, dx^i\, dx^i\,,\label{ghmet}
\ee
where $V$ and $A_i$ depend only on the $x^i$.  In the orthonormal
basis
\be
e^0 = V^{-1/2}\, (d\tau + A_i\, dx^i)\,,\qquad e^i = V^{1/2}\, dx^i\,,
\ee
the spin connection is then given by
\be
\omega_{0i} = \ft12 V^{-3/2}\, [-\del_i\, V\, e^0 + F_{ij}\,
e^j]\,,\qquad
\omega_{ij} = \ft12 V^{-3/2}\,[ \del_j V\, e^i - \del_i V\, e^j -
F_{ij}\, e^0]\,,\label{ghspincon}
\ee
where we have defined $F_{ij} \equiv \del_i\, A_j -\del_j\, A_i$.  It
is convenient to introduce a ``dual'' $\wtd\omega_{0i}$ of the
spin-connection components $\omega_{ij}$, as $\wtd\omega_{0i} \equiv
\ft12\ep_{ijk}\, \omega_{jk}$.  It is easily seen that if the spin
connection is self-dual or anti-self-dual, in the sense that
$\wtd\omega_{0i} = \pm \omega_{0i}$, then the curvature 2-forms
$\Theta$ are self-dual or anti-self-dual, not only in the analogous
sense $\Theta_{0i} = \pm\ft12 \ep_{ijk}\, \Theta_{jk}$, but also in
the normal sense $\Theta_{ab} = \pm {*\Theta_{ab}}$.  In particular,
when this condition is satisfied, the metric is Ricci flat.

  It is easy to see from (\ref{ghspincon}) that the spin-connection
satisfies $\wtd\omega_{0i} = \pm \omega_{0i}$ if and only if the
metric functions satisfy
\be
\vec\nabla\, V = \pm \vec\nabla\times\vec A\,,\label{gheq}
\ee
where the expressions here are the standard ones of three-dimensional
Cartesian coordinates.  In other words, $\del_i\, V = \pm \ep_{ijk}\,
\del_j\, A_k$.  Thus (\ref{gheq}) is the condition for (\ref{ghmet})
to be Ricci flat, and furthermore, self-dual or anti-self-dual.  In
particular, taking the divergence of (\ref{gheq}) we get
$\nabla^2\, V=0$, 
so $V$ should be harmonic, and then $\vec A$ can be solved for (modulo
a gauge transformation) using (\ref{gheq}).

   The multi-centre instantons \cite{gibhaw} are obtained by taking
\be
V = c + \sum_\a \fft{q_\a}{|\vec x - \vec x_\a|}\,,
\ee
where $c$ and $q_\a$ are constants.  If $c=0$ one gets the multi
Taub-NUT metrics, while if $c\ne0$ (conveniently one chooses $c=1$),
the metrics are instead multi Eguchi-Hanson.

    If we take a uniform distribution of charges spread over a
two-dimensional plane of radius $R$, then at a perpendicular distance
$y$ from the centre of the disc the potential $V$ is given by
\bea
V &=& c  + q\, \int_0^R \fft{r\, dr}{(y^2 + r^2)^{1/2}}\,,\nn\\
&=& c +  q\, \Big[ (R^2+y^2)^{1/2} -|y|\, \Big]\,,\\
&=& c+ q\, R - q\, |y| + {\cal O}(1/R)\,.\nn
\eea
Thus if we send $R$ to infinity, while setting $q=-m$ and adjusting
$c$ such that $c+ q\, R=1$, we obtain
\be
V = 1 + m\, |y|\,.
\ee
It is easy to establish from (\ref{gheq}) that a solution for $\vec A$
is then
\be
\vec A=  (0, m\, z_3, 0)\,,
\ee
(where we take the Cartesian coordinates to be $\vec x = (y, z_2,
z_3)$), and so we have arrived back at our original metric
(\ref{4met}).  It can therefore be described as a continuum of
Taub-NUT instantons distributed uniformly over a two-dimensional
plane.  (This is essentially the construction of \cite{berg78dual}.)  

    A more physical picture of this limit is that there can be
multi-instanton generalisations of an AC manifold, and a uniform
distribution would turn the non-abelian isometry group into an abelian
$U(1)$ group.

\section{Orientifold planes and the Atiyah-Hitchin
metric}\label{atiyahsec}

    In addition to D-branes, which have positive tension, string
theory admits orientifold planes which have negative tension. Since
orientifold planes are not dynamical, the negative tension does not
lead to instabilities as it would if the tension of an ordinary
D-brane were negative. This is because they are pinned in position:
the inversion symmetry employed in the orientifold projection excludes
translational zero modes.

    In M-theory, an orientifold plane corresponds to the product of
the Atiyah-Hitchin metric \cite{ah,gibman} with seven-dimensional
Minkowski space-time \cite{orientifolds}. The Atiyah-Hitchin metric is
a smooth nonsingular hyper-K\"ahler 4-metric and hence BPS. It is
invariant under $SO(3)$ acting on principal orbits of the form
$SO(3)/\Z_2$, where the $\Z_2$ is realised as the group of diagonal
$SO(3)$ matrices. Near infinity, it is given approximately by the
Taub-NUT metric divided by CP, taken with a {\it negative} ADM
mass. The CP quotient symmetry here takes $(\psi,x^i)$ to
$(-\psi,-x^i)$, where $\psi$ is the Kaluza-Klein coordinate.  As in
string theories, the negative mass does not lead to instabilities
because this quotient symmetry of the asymptotic metric eliminates
translational zero modes.

   One might think that being BPS and having negative ADM mass would be
inconsistent because of the Positive Mass Theorem.  However the
Positive Mass Theorem for ALF spaces such as the Atiyah-Hitchin metric
is rather subtle \cite{mass}.  Suffice it to say that one needs to
solve the Dirac equation subject to boundary conditions at infinity as
an essential ingredient in the proof. If the manifold is simply
connected there is a unique spin structure but because a neighbourhood
of infinity where one imposes the boundary conditions is not
simply-connected, it is not obvious that a suitable global solution
exists in the unique spin structure .  In the Atiyah-Hitchin case it
seems clear that it does not.
 
    This example shows that in principle, gravity can resolve
singularities associated with branes of negative tension. However, to
make this more precise we would need to consider the unresolved
spacetime and its relationship to the Atiyah-Hitchin metric. To lowest
order, one  might consider this to be  the flat metric on $S^1\times \R^3$ with
coordinates $(\psi, {\bf x})$. This clearly has a singularity
at $(0,0,0,0)$. However this approximation ignores the
Kaluza-Klein magnetic field generated by the orientifold plane.
If we maintain spherical symmetry we would be led at the next level of
approximation to the Taub-NUT metric with negative mass:
\be
ds^2 = 4 ( 1+ {2M \over r} ) ^{-1} (d \psi +\cos \theta d \phi))
+ ( 1+ {2M \over r} ) ( dr ^2 + r^2 ( \theta ^2 + \sin \theta d \phi ^2 ).
\ee  
$CP$ acts as $(\psi, \theta, \phi) \rightarrow ( -\pi, \pi-\theta, \phi
+ \pi)$ and the ADM mass $M$ is negative. Clearly the Taub-NUT
approximation breaks down at small positive  $r$ because if $r< -2M$,
the metric signature is $----$ rather than $ ++++$.

The full non-singular Atiyah-Hitchin metric can be written as
\be
ds^2 = dt^2  + a^2(t) \sigma_1 ^2 + b^2(t)  \sigma_2 ^2 
+ c^2(t)\sigma_3^2   ,
\ee
where $\sigma _1, \sigma _2, \sigma _3$ are Cartan-Maurer forms for
$SU(2)$ and the allowed range of angular coordinates is restricted 
by the fact that $CP$ should act as the identity.
For large $t$ we have $a \rightarrow b$ and the metric 
tends to the Taub-NUT metric. 

The Atiyah-Hitchin metric and Taub-NUT metric are members of a general
family of locally $SU(2)$-invariant hyper-K\"ahler metrics in which
the three K\"ahler forms transform as a triplet. They satisfy a set of
first order differential equations coming from a superpotential as
described in section \ref{superpotentialsec}. The Eguchi-Hanson metric
discussed above is a member of another family of locally
$SU(2)$-invariant hyper-K\"ahler metrics in which the three K\"ahler
forms transform as singlets \footnote{One says that $SU(2)$ acts
triholomorphically in this case}. They satisfy a different set of
first order differential equations also given in section
\ref{superpotentialsec}.  One may check that the Heisenberg limits of
these two sets of equations are identical and the solutions are
precisely the metrics (\ref{4met}) in which the Heisenberg group acts
triholomorphically. Thus in the Heisenberg limit the triplet becomes
three singlets.

One may also take the Heisenberg limit directly in the asymptotic
Taub-NUT metric. This also leads to the metric (\ref{4met}).  However
in order to build in the projection under CP one may identify $(z_1,
z_2, z_3)$ with $(-z_1, -z_2, z_3)$.  Thus by analogy with the
construction of section \ref{stacksec} we propose that the singular
unresolved metric for a stack of orientifold planes analogous to the
metric of a stack of D6-branes is (\ref{4met}) with this additional
identification.

\section{Domain-walls from pure gravity}\label{dwsec}

    The eight-dimensional domain-wall example of the previous section
lifted to a solution of eleven-dimensional supergravity with vanishing
CJS field $F_\4$.  Its transverse coordinate $y$, together with the
three toroidal coordinates $(z_1,z_2,z_3)$ of the reduction from
$D=11$, gave the four coordinates of the cohomogeneity one Ricci-flat
Heisenberg metric, which could be viewed as a limit of the
Eguchi-Hanson metric.  The principal orbits in the Heisenberg limit
were $T^1$ bundles over $T^2$.

   In this section we shall generalise this construction by
considering domain-wall solutions in maximal supergravities that lift
to give purely geometrical solutions in eleven dimensions.  Thus if we
begin with such a domain-wall in $D$-dimensional supergravity, the
metric after lifting to eleven dimensions will be $ds_{11}^2 =
dx^\mu\, dx_\mu + ds_{12-D}^2$, where $ds_{12-D}^2$ is a Ricci-flat
metric of cohomogeneity one, with principal orbits that are again of
the form of torus bundles.

   The cases that we shall consider here arise from domain-wall
solutions in $D=7$, 6 and 5. Correspondingly, these give rise to
Ricci-flat Heisenberg metrics of dimensions 6, 7 and 8.  Since each
domain wall preserves a fraction of the supersymmetry, it follows that
the associated Ricci-flat metrics admit certain numbers of
covariantly-constant spinors.  In other words, they are metrics of
special holonomy.  This property generalises the special holonomy of
the 4-dimensional Ricci-flat K\"ahler metric in the previous example
in section \ref{egsec}.  Specifically, for the Ricci-flat metrics in
dimension 6, 7 and 8 we shall see that the special holonomies $SU(3)$,
$G_2$ and Spin(7) arise.

\subsection{Six-dimensional manifolds with $SU(3)$ holonomy}

\subsubsection{$T^1$ bundle over $T^4$}\label{t1overt4ssec}

   In this first six-dimensional example, the principal orbits form an
$T^1$ bundle over $T^4$.  It arises if we take a domain wall solution in
six spacetime dimensions, supported by the two 0-form field strengths
$\cF^1_{\0 23}$ and $\cF^1_{\0 45}$, carrying equal charges $m$.  In
what follows, we shall adhere to the terminology ``charges,'' although
in some circumstances it may be more appropriate to think of
``fluxes.''  The domain-wall metric is
\be
ds^2 = H^{1/2}\, dx^\mu\, dx_\mu + H^{5/2}\, dy^2\,,\label{d6dw}
\ee
where $H=1+ m\, |y|$.  Oxidising back to $D=11$, we get the
eleven-dimensional metric
\be
d\hat s^2 =  dx^\mu\, dx_\mu + H^{-2}\, [dz_1 + m\, (z_3\, dz_2 +
z_5\, dz_4)]^2 + H^2\, dy^2  + H\, (dz_2^2 + \cdots + dz_5^2)\,.
\ee
Thus we conclude that the six-dimensional metric
\be
ds_6^2= H^{-2}\, [dz_1 + m\, (z_3\, dz_2 +
z_5\, dz_4)]^2 + H^2\, dy^2  + H\, (dz_2^2 + \cdots +
 dz_5^2)\label{6met}
\ee
is Ricci flat.  Since the solution carries two charges it preserves
$\ft14$ of the supersymmetry, and so this 6-metric must have $SU(3)$
holonomy.  Thus it is a Ricci-flat K\"ahler 6-metric.
 
    Define an orthonormal basis by
\bea
&&e^0 = H\, dy\,,\qquad
e^1 = H^{-1}\, [dz_1 + m\, (z_3\, dz_2 + z_5\, dz_4)] \,,\nn\\
&& e^2= H^{1/2}\, dz_2\,,\qquad e^3 = H^{1/2}\, dz_3\,,\qquad
   e^4= H^{1/2}\, dz_4\,,\qquad e^5 = H^{1/2}\, dz_5\,.
\eea
It can then be seen that the K\"ahler form is given by
\be
J = e^0\wedge e^1 -e^2\wedge e^3 -e^4\wedge e^5\,.\label{62form}
\ee

     Following the same strategy as in the previous section, we can
obtain the differential equations whose solutions define complex
coordinates $\zeta^\mu$ in terms of the real coordinates.  First,
define a new real coordinate $x$ in place of $y$, such that $H^2\, dy
\equiv dx$, and hence $y+ m\, y^2 + \ft13 m^2\, y^3 = x$.  This
implies that $H=(1+3m\, x)^{1/3}$. After straightforward algebra, we
find that a suitable choice for the definition of the complex
coordinates is
\bea
&&\zeta_1 = z_3 + \im\, z_2\,,\qquad \zeta_2 = z_5 + \im\, z_4\,,\nn\\
&&\zeta_3 = x+ \im\, z_1 + \ft14 m\, (z_2^2+z_4^2 -3z_3^2 -3 z_5^2
 -2\im\, z_2\, z_3 -2\im\, z_4\, z_5)\,.
\eea
The metric then takes the form
\be
ds_6^2 = H\, (|d\zeta_1|^2 + |d\zeta_2|^2 ) + H^{-2}\, |d\zeta_3 +
\ft12 m \,(\bar\zeta_1\, d\zeta_1 + \bar\zeta_2\, d\zeta_2)|^2\,,
\label{6met4}
\ee
the harmonic function $H$ is given by
\be
H= \Big[1+ \ft32m\, (\zeta_3+\bar\zeta_3) +
        \ft34 m^2\,(|\zeta_1|^2 + |\zeta_2|^2)\Big]^{1/3}\,,\label{kah4}
\ee
and the K\"ahler function is $K=H^4/m^2$.

    In the case that $H=m\, y$, there is a scaling invariance
of the metric (\ref{6met}) generated by the homothetic Killing vector
\be
D= y \, \fft{\del}{\del y} + 3 z_1\, \fft{\del }{\del z_1} 
+ \ft32 z_2\, \fft{\del }{\del z_2} 
+ \ft32 z_3\, \fft{\del }{\del z_3} 
+ \ft32 z_4\, \fft{\del }{\del z_4} 
+ \ft32 z_5\, \fft{\del }{\del z_5} \,.
\ee
In addition,  (\ref{6met}) is invariant under the linear action of $U(2)$ 
on $(z_2,z_3,z_4,z_5)$ preserving the self-dual 2-form $dz_3\wedge
dz_2 + dz_5\wedge dz_4$.

\subsubsection{$T^2$ bundle over $T^3$}
 
  There is a second type of 2-charge domain wall in $D=6$, again
supported by two 0-form field strengths coming from the Kaluza-Klein
reduction of the eleven-dimensional metric.  A representative example
is given by using the field strengths $(\cF_{\0 34}^1, \cF_{\0
35}^2)$.  The domain-wall metric is again given by (\ref{d6dw}), but
now, upon oxidation back to $D=11$, we obtain $d\hat s^2 = dx^\mu\,
dx_\mu + ds_6^2$ with the Ricci-flat 6-metric now given by
\bea
ds_6^2 &=& H^2\, dy^2 + H^{-1}\, (dz_1 + m\, z_4\, dz_3)^2 + H^{-1}\,
(dz_2 + m\, z_5\, dz_3)^2 \nn\\
&&+ H^2\, dz_3^2 + H\, (dz_4^2 + dz_5^2)\,.\label{cy2met}
\eea
Defining the orthonormal basis
\bea
&&e^0 = H\, dy\,,\qquad e^1 = H^{-1/2}\, (dz_1 + m\, z_4\,
dz_3)\,,\qquad e^2 = H^{-1/2}\, (dz_2 + m\, z_5\, dz_3)\,,\nn\\
&& e^3 = H\, dz_3\,,\qquad e^4 = H^{1/2}\, dz_4\,,\qquad e^5=H^{1/2}\,
dz_5\,,
\eea
we find that the torsion-free spin connection is given by
\bea
&&\omega_{01} = \lambda\, e^1\,,\qquad \omega_{02} = \lambda\,
e^2\,,\qquad \omega_{03}=-2\lambda\, e^3\,,\nn\\
&&\omega_{04} = -\lambda\, e^4\,,\qquad \omega_{05}=-\lambda\,
e^5\,,\qquad \omega_{12}=0\,,\nn\\
&&\omega_{13} = -\lambda\, e^4\,,\qquad \omega_{14}= \lambda\,
e^3\,,\qquad \omega_{15}=0\,,\\
&&\omega_{23}= -\lambda\, e^5\,,\qquad \omega_{24}=0\,,\qquad
\omega_{25}= \lambda\, e^3\,,\nn\\
&&\omega_{34}= \lambda\, e^1\,,\qquad \omega_{35} = \lambda\,
e^2\,,\qquad \omega_{45} =0\,,\nn
\eea
where $\lambda\equiv \ft12 m\, H^{-2}$.

    From this, it is easily established that the following 2-form is
covariantly constant:
\be
J = e^0\wedge e^3 + e^1\wedge e^4 + e^2\wedge e^5\,.\label{cy2j}
\ee
This is the K\"ahler form.  From this, using the same strategy as we
used in previous sections, we can deduce that the following are a
suitable set of complex coordinates:
\be
\zeta_1 = z_1 + \im\, H\, z_4\,,\qquad \zeta_2 = z_2 + \im\, H\,
z_5\,,\qquad \zeta_3 = y + \im\, z_3\,.
\ee
The Hermitean metric tensor $g_{\mu\bar \nu}$ can then be derived from
the K\"ahler function
\be
K = -\fft1{4H}\, \Big[ (\zeta_1-\bar\zeta_1)^2 +
(\zeta_2-\bar\zeta_2)^2\Big] + \ft1{48}\, |\zeta_3|^2\, \Big[ 8(H^2
+H+1) - m^2\, |\zeta_3|^2\Big]\,.
\ee

    In the case that $H=m\, y$, there is a scaling invariance
of the metric (\ref{cy2met}) generated by the homothetic Killing vector
\be
D= y \, \fft{\del}{\del y} + \ft52 z_1\, \fft{\del }{\del z_1} 
+ \ft52 z_2\, \fft{\del }{\del z_2} 
+  z_3\, \fft{\del }{\del z_3} 
+ \ft32 z_4\, \fft{\del }{\del z_4} 
+ \ft32 z_5\, \fft{\del }{\del z_5} \,.
\ee
In addition, (\ref{cy2met}) is invariant under the linear action of
$SO(2)$ on $(z_1,z_2)$ and $(z_4,z_5)$.

\subsection{Seven-dimensional manifolds with $G_2$ holonomy}

\subsubsection{$T^2$ bundle over $T^4$}

   Now consider a 4-charge domain wall in $D=5$.  Take the charges to
be carried by the following 0-form field strengths: $(\cF_{\0 34}^1,
\cF_{\0 56}^1, \cF_{\0 35}^2, \cF_{\0 46}^2)$.  Note that here, unlike
the 2-charge cases in $D=6$, it matters what the
relative signs of the charges are here, in order to get a
supersymmetric solution.  Specifically, the Bogomolny'i matrix ${\cal
M}$ is given here by \cite{lpsol}
\be
{\cal M} = \mu\, \oneone + q_1\, \Gamma_{y 134} + q_2\, \Gamma_{y 1 56} +
q_3 \, \Gamma_{y 2 35} + q_4\, \Gamma_{y 2 46}\,,
\ee
where
\be
\mu = \sum_i |q_i|\,.
\ee
Having fixed a set of conventions, the signs of the first three
charges $(q_1, q_2,q_3)$ can be arbitrary for supersymmetry, but only
for one sign of the fourth charge $q_4$ is there supersymmetry.  With
our conventions, it must be negative.  Thus we may consider:
\be
q_1=q_2=q_3=-q_4=m\,.
\ee
 
    The domain-wall metric in $D=5$ is
\be
ds^2 = H^{4/3}\, dx^\mu\, dx_\mu + H^{16/3}\, dy^2\,.
\ee
Oxidising back to $D=11$ in the standard way, we get $d\hat
s^2=dx^\mu\, dx_\mu + ds_7^2$, where the seven-dimensional metric is
given by
\bea
ds_7^2 &=& H^4\, dy^2 + H^{-2}\, [dz_1 + m\, (z_4\, dz_3 + z_6\,
dz_5)]^2 + H^{-2}\, [dz_2 + m\, (z_5\, dz_3 - z_6\, dz_4)]^2 \nn\\
&&+H^2\, (dz_3^2 + \cdots + dz_6^2)\,.\label{7met}
\eea
Note that the minus sign in the term involving $-z_6\, dz_5$ is a
reflection of the fact that the charge associated with $\cF_{\0 46}^2$
is negative for supersymmetry.  On account of the supersymmetry, we
conclude that the Ricci-flat metric $ds_7^2$ admits one
covariantly-constant spinor, and thus it must have $G_2$ holonomy.
 
    Define the orthonormal basis
\bea
&&e^0=H^2\, dy\,,\qquad
e^3= H\, dz_3\,,\qquad e^4 = H\, dz_4\,,\qquad e^5=H\, dz_5\,,\qquad
e^6=H\, dz_6\,,\nn\\
&&e^1 = H^{-1}\, [dz_1 + m\, (z_4\, dz_3 + z_6\,
dz_5)]\,,\qquad e^2 = H^{-1}\, [dz_2 + m\, (z_5\, dz_3 - z_6\,
dz_4)]\,.
\eea
In this basis, the torsion-free spin connection is given by
\bea
&&\omega_{01} = 2\lambda\, e^1\,,\qquad \omega_{02} = 2\lambda\,
e^2\,,\qquad
\omega_{03} = -2\lambda\, e^3\,,\nn\\
&&\omega_{04} = -2\lambda\, e^4\,,\qquad \omega_{05} = -2\lambda\,
e^2\,,\qquad
\omega_{06} = -2\lambda\, e^6\,,\nn\\
&&\omega_{12}=0\,,\qquad \omega_{13} = -\lambda\, e^4\,,\qquad
\omega_{14} = \lambda\, e^3\,,\nn\\
&&\omega_{15}=-\lambda\, e^6\,,\qquad \omega_{16}= \lambda\,
e^5\,,\qquad \omega_{23} =-\lambda\, e^5\,,\nn\\
&&\omega_{24} = \lambda\, e^6\,,\qquad \omega_{25} = \lambda\,
e^3\,,\qquad \omega_{26}= -\lambda\, e^4\,,\\
&&\omega_{34} =\lambda\, e^1\,,\qquad \omega_{35} = \lambda\,
e^2\,,\qquad \omega_{36}=0\,,\nn\\
&&\omega_{45}= 0\,,\qquad \omega_{46} = -\lambda\, e^2\,,\qquad
\omega_{56} = \lambda\, e^1\,,\nn
\eea
where $\lambda\equiv \ft12 m\, H^{-3}$.
 
     It can now be verified that the following 3-form is covariantly
constant:
\bea
\psi_\3 &\equiv& e^0\wedge e^1\wedge e^2 - e^1\wedge e^4\wedge e^6 +
e^1\wedge e^3\wedge e^5 \nn\\
&&- e^2\wedge e^5\wedge e^6 -e^2\wedge e^3\wedge e^4 - e^0\wedge
e^3\wedge ^6 - e^0\wedge e^4\wedge e^5\,.
\eea
The existence of such a 3-form is characteristic of a 7-manifold with
$G_2$ holonomy.  In fact the components $\psi_{ijk}$ are the structure
constants of the multiplication table of the seven imaginary unit
octonions $\gamma_i$:
\be
\gamma_i\, \gamma_j = -\delta_{ij} + \psi_{ijk}\, \gamma_k\,.
\ee
Note that if the sign of the $z_6\, dz_4$ term in (\ref{7met}) had
been taken to be $+$ instead of $-$ (while keeping all other
conventions unchanged), then there would not exist a
covariantly-constant 3-form.  This is another reflection of the fact
that the occurrence of supersymmetry is dependent up on sign of the
fourth charge.  (At the same time as the 8-dimensional spinor
representation of the $SO(7)$ tangent-space group decomposes as
$8\longrightarrow 7+1$ under $G_2$, the 35-dimensional antisymmetric
3-index representation decomposes as $35\longrightarrow 27+7+1$.  It is
the singlet in each case that corresponds to the covariantly-constant
spinor (Killing spinor) and 3-form.)
 
    Note that we can write the 3-form $\psi_\3$ as
\be
\psi_\3 = e^0\wedge e^1\wedge e^2
-e^0\wedge K_0 - e^1\wedge K_1 - e^2\wedge K_2 \,,
\ee
where
\be
K_0\equiv e^3\wedge e^6 + e^4\wedge e^5 \,,\qquad
K_1\equiv e^4\wedge e^6 - e^3\wedge e^5\,,\qquad
K_2\equiv e^5\wedge e^6 + e^3\wedge e^4\,.
\ee
The three 2-forms $K_0$, $K_1$ and $K_2$ are self-dual with respect to
the metric in the $(3,4,5,6)$ directions.  Thus in the entire
construction, both of the 7-dimensional metric $ds_7^2$ and the
covariantly-constant 3-form $\psi_\3$, the flat 4-torus metric $dz_3^2
+ \cdots +dz_6^2$ can be replaced by any hyper-K\"ahler 4-metric.
(The potential terms that ``twist'' the fibres in the $z_1$ and $z_2$
directions are now replaced by potentials for the self-dual 2-forms
$K_2$ and $K_1$.  See section \ref{g2sec} below.)

    In the case that $H=m\, y$, there is a scaling invariance
of the metric (\ref{7met}) generated by the homothetic Killing vector
\be
D= y \, \fft{\del}{\del y} + 4 z_1\, \fft{\del }{\del z_1} 
+ 4 z_2\, \fft{\del }{\del z_2} 
+ 2 z_3\, \fft{\del }{\del z_3} 
+ 2 z_4\, \fft{\del }{\del z_4} 
+ 2 z_5\, \fft{\del }{\del z_5} 
+ 2 z_6\, \fft{\del }{\del z_6} \,.
\ee
In addition, (\ref{7met}) is invariant under the linear action of
$SU(2)$ on $(z_3,z_4,z_5,z_6)$ that preserves the two 2-forms
$dz_4\wedge dz_3 + dz_6\wedge dz_5$ and $dz_5\wedge dz_3 -dz_6\wedge
dz_4$.  The 6-dimensional nilpotent algebra in this case is the
complexification of the standard 3-dimensional $\frak{ur}$-Heisenberg
algebra.

\subsubsection{$T^3$ bundle over $T^3$}

    There is an inequivalent class of domain-wall solutions
in five-dimensional spacetime, for which a representative example 
is supported by the three fields
$\{\cF^1_{\0 56}, \cF^2_{\0 46}, \cF^3_{\0 45} \}$.   This gives
the Ricci-flat 7-metric
\bea
ds_7^2 &=& H^3\, dy^2 + H^{-1}\,(dz_1+m\, z_6\, dz_5)^2 + H^{-1}\, (dz_2
 -m\, z_6\, dz_4)^2 \nn\\
&&+ H^{-1}\, (dz_3 + m\, z_5\, dz_4)^2
+ H^2 (dz_4^2 + dz_5^2 + dz_6^2)\,.\label{7met2}
\eea
In the obvious orthonormal basis $e^0=H^{3/2}\, dy$, $e^2=H^{-1/2}\,
(dz_1 + m\, z_6\, dz_5)$, {\it etc}., the covariantly-constant
associative 3-form is given by
\bea
\psi_\3 &=& e^0\wedge e^1\wedge e^4 + e^0\wedge e^2\wedge e^5 + 
   e^0\wedge e^3\wedge e^6\nn\\
&&+ e^1\wedge e^2\wedge e^6 + e^2\wedge e^3\wedge e^4 + e^3\wedge
e^1\wedge e^5 - e^4\wedge e^5\wedge e^6\,.
\eea

    In the case that $H=m\, y$, there is a scaling invariance
of the metric (\ref{7met2}) generated by the homothetic Killing vector
\be
D= y \, \fft{\del}{\del y} + 3 z_1\, \fft{\del }{\del z_1} 
+ 3 z_2\, \fft{\del }{\del z_2} 
+ 3 z_3\, \fft{\del }{\del z_3} 
+ \ft32 z_4\, \fft{\del }{\del z_4} 
+ \ft32 z_5\, \fft{\del }{\del z_5} 
+ \ft32 z_6\, \fft{\del }{\del z_6} \,.
\ee
In addition, (\ref{7met2}) is invariant under
$SO(3)$ acting linearly on $(z_1,z_2,z_3)$ and $(z_3,z_4,z_5,z_6)$.  

\subsection{Eight-dimensional manifolds with Spin(7) holonomy}

\subsubsection{$T^3$ bundle over $T^4$}

    Now consider a 6-charge domain wall solution in $D=4$, supported
by the 0-form field strengths $(\cF_{\0 45}^1, \cF_{\0 67}^1, \cF_{\0
46}^2, \cF_{\0 57}^2, \cF_{\0 47}^3, \cF_{\0 56}^3)$.  As in the
previous case, the signs of the charges must be appropriately chosen
in order to have a supersymmetric solution.  In $D=4$, the domain wall
metric, with all charges chosen equal in magnitude, is
\be
ds^2 = H^3\, dx^\mu\, dx_\mu + H^9\, dy^2\,.\label{above8}
\ee
Oxidising back to $D=11$ gives the eleven-dimensional metric $d\hat
s^2= dx^\mu\, dx_\mu + ds_8^2$, where the Ricci-flat 8-metric is given
by
\bea
ds_8^2 &=& H^6\, dy^2 + H^{-2}\, [dz_1 +m\, (z_5\, dz_4 + z_7\, dz_6)]^2
+ H^{-2}\, [dz_2 +m\, (z_6\, dz_4 - z_7\, dz_5)]^2 \nn\\
&&+ H^{-2}\, [dz_3 +m\, (z_7\, dz_4 + z_6\, dz_5)]^2
+ H^3\, (dz_4^2 + \cdots + dz_7^2)\,.\label{d8metric}
\eea
Since the solution preserves $1\ft1{16}$ of the supersymmetry, it follows
that this Ricci-flat 8-metric must have Spin(7) holonomy.

   Let us choose the natural orthonormal basis,
\bea
&&e^0=H^3\, dy\,,\qquad
e^1 = H^{-1}\, [dz_1 + m\, (z_5\, dz_4 + z_7\,
dz_6)]\,,\nn\\
&&
e^2 = H^{-1}\, [dz_2 + m\, (z_6\, dz_4 - z_7\, dz_5)]\,,\qquad
e^3 = H^{-1}\, [dz_3 + m\, (z_7\, dz_4 + z_6\, dz_5)]\,,\nn\\
&&e^4= H^{3/2}\, \, dz_4\,,\qquad e^5 = H^{3/2}\, dz_5\,,
\qquad e^6=H^{3/2}\, dz_6\,,\qquad
e^7=H^{3/2}\, dz_7\,.
\eea
The spin connection is then given by $\omega_{ij} = \ft12(c_{ijk} +
c_{ikj} - c_{kji})\, e^k$, where the non-vanishing connection
coefficients $c_{ijk}= - c_{jik}$ are specified by
\bea
&&c_{01}{}^1 = c_{45}{}^1 =c_{67}{}^1 = c_{02}{}^2=c_{46}{}^2=-c_{57}{}^2
=c_{03}{}^3=c_{47}{}^3 =c_{56}{}^3=2\lambda\,,\\
&&c_{04}{}^4 =c_{05}{}^5=c_{06}{}^6=c_{07}{}^7 = -3\lambda\,,\nn
\eea
and $\lambda \equiv \ft12 m\, H^{-4}$ here.
 
    From this, it is straightforward to show that the following 4-form
is covariantly constant:
\bea
\Psi_\4&=& \!\! -(e^0\wedge e^1 + e^2\wedge e^3)\wedge (e^4\wedge e^5 +
e^6\wedge e^7)
-(e^0\wedge e^2 - e^1\wedge e^3)\wedge (e^4\wedge e^6 -
e^5\wedge e^7) \nn\\
&&\!\!\!-(e^0\wedge e^3 + e^1\wedge e^2)\wedge (e^4\wedge e^7 +
e^5\wedge e^6) + e^0\wedge e^1\wedge e^2\wedge e^3 + e^4\wedge
e^5\wedge e^6\wedge e^7\,.\label{4form}
\eea
The existence of this 4-form, which is self-dual, is characteristic
of 8-manifolds with Spin(7) holonomy.

    In the case that $H=m\, y$, there is a scaling invariance
of the metric (\ref{d8metric}) generated by the homothetic Killing vector
\be
D= y \, \fft{\del}{\del y} + 5 z_1\, \fft{\del }{\del z_1} 
+ 5 z_2\, \fft{\del }{\del z_2} 
+ 5 z_3\, \fft{\del }{\del z_3} 
+ \ft52 z_4\, \fft{\del }{\del z_4} 
+ \ft52 z_5\, \fft{\del }{\del z_5} 
+ \ft52 z_6\, \fft{\del }{\del z_6} 
+ \ft52 z_7\, \fft{\del }{\del z_7} \,.
\ee
In addition, (\ref{d8metric}) is invariant under the linear action of
$SU(2)$ on $(z_4,z_5,z_6,z_7)$ that preserves the three 2-forms
$dz_5\wedge dz_4 + dz_7\wedge dz_6$, $dz_6\wedge dz_4 -dz_7\wedge
dz_5$ and $dz_7\wedge dz_4 + dz_6\wedge dz_5$.  

\subsection{Further examples and specialisations}

    The various domain walls that we have obtained above are the most
natural ones to consider, since they possess the maximum number of
charges in each case, and after lifting to $D=11$ they are
irreducible.  It is, nevertheless, of interest also to study some of
the other possible examples.

\subsubsection{$T^1$ bundle over $T^3$}\label{t1overt3ssec}

   In seven-dimensional maximal supergravity, the largest number of
allowed charges for domain walls is 1 (see, for example,
\cite{classp}).  The metric is given by
\be
ds_7^2 = H^{1/5}\, dx^\mu\, dx_\mu + H^{6/5}\, dy^2\,,
\ee
where $H=1+m\, |y|$.  After lifting to $D=11$, this gives $ds_{11}^2 =
dx^\mu\, dx_\mu + ds_5^2$, where $ds_5^2$ is the Ricci-flat metric
\be
ds_5^2 = H\, dy^2 + H^{-1}\, (dz_1 + m\, z_3\, dz_2)^2 + H\, (dz_2^2 +
dz_3^2) + dz_4^2\,.
\ee
This is clearly reducible, being nothing but the direct sum of the
four-dimensional Ricci-flat metric (\ref{4met}) and a circle.  It is
for this reason that we omitted this 5-dimensional example in our
enumeration above.  It can be viewed as a $T^1$ bundle over $T^3$, but
since the bundle is trivial over a $T^1$ factor in the base, it would
be more accurate to describe it as $T^1$ times a $T^1$ bundle over
$T^2$.

\subsubsection{Ricci-flat metrics with fewer charges}

    There are many possibilities for obtaining other Ricci-flat
metrics, by turning on only subsets of the charges in the metrics we
have already obtained.  We shall illustrate this by considering the
example of the 8-dimensional metric (\ref{d8metric}).  If we introduce
parameters $\ep_i$, where $\ep_i=1$ if the $i$'th of the six charges
listed above (\ref{above8}) is turned on, an $\ep_i=0$ if the $i$'th
charge is turned off.  After lifting the resulting domain wall from
$D=4$ to $D=11$, we get a Ricci-flat 8-metric given by
\bea
ds_8^2 &=& H^{\sum_i \ep_i}\, dy^2 + H^{-\ep_1-\ep_2}\, h_1^2 
+ H^{-\ep_3-\ep_4}\, h_2^2 + H^{-\ep_5-\ep_6}\, h_3^2\nn\\
&& +H^{\ep_1+\ep_3+\ep_5}\, h_4^2 + 
   H^{\ep_1+\ep_4+\ep_6}\, h_5^2 + 
   H^{\ep_2+\ep_3+\ep_6}\, h_6^2 + 
   H^{\ep_2+\ep_4+\ep_5}\, h_7^2 \,,
\eea
\bea
&&h_1 = dz_1 + \ep_1\, z_5\, dz_4 + \ep_2\, z_7\, dz_6\,,\quad
h_2 = dz_2 + \ep_3\, z_6\, dz_4 - \ep_4\, z_7\, dz_5\,,\\
&&h_3 = dz_3 + \ep_5\, z_7\, dz_4 + \ep_6\, z_6\, dz_5\,,\quad
h_4 = dz_4\,,\quad h_5=dz_5\,,\quad h_6=dz_6\,,\quad
 h_7 =dz_7\,.\nn
\eea

\subsubsection{$SU(4)$ holonomy in $D=8$}

   There are other possibilities, which involve a lesser number of
charges which are not themselves a subset of the maximal set.  For
example, we can consider the following 8-dimensional Ricci-flat metric
that comes from lifting a 3-charge four-dimensional domain wall,
supported by the fields $\cF^1_{\0 23}$, $\cF^1_{\0 45}$, $\cF^1_{\0
67}$.  This gives
\be
ds_8^2 = H^3\, dy^2 + H^{-3}\, (dz_1 + z_3\, dz_2 + z_5\, dz_4 + z_7\,
dz_6)^2 + H\, (dz_2^2 + \cdots +dz_7^2)\,.
\ee
This metric has $SU(4)$ holonomy, and it can be viewed as a Heisenberg
limit of a complex line bundle over a six-dimensional
Einstein-K\"ahler space such as $S^2\times S^2 \times S^2$, or $\CP^3$.

\section{Heisenberg limits of complete metrics of special holonomy}

   In this section, we generalise the discussion of the Heisenberg
limit of the Eguchi-Hanson metric that we gave in section
\ref{egheissec}, and show how the various Ricci-flat metrics that we
obtained from domain-wall solutions in section \ref{dwsec} can be
viewed as arising as Heisenberg limits of complete metrics of special
holonomy.

\subsection{Contractions of Ricci-flat K\"ahler 6-metrics}

\subsubsection{Contractions of $T^1$ bundles over Einstein-K\"ahler}

   The contraction to the Heisenberg limit of the Eguchi-Hanson metric
was discussed in section \ref{egheissec} at the level of the
metric itself, and in section \ref{superpotentialsec} at the level of
the equations of motion and superpotential.  This contraction
procedure can be easily generalised to higher dimensions.  In
particular, we may obtain the six-dimensional Ricci-flat Heisenberg
metric (\ref{6met}) as a contraction of a Ricci-flat metric on a line
bundle over an Einstein-K\"ahler 4-metric with positive scalar
curvature, such as $\CP^2$.  If we consider the more general case of a
line bundle over $\CP^n$, the starting point will be the metric
\be
ds_{2n+2}^2 = dt^2 + a^2\, \sigma^\a\, \bar \sigma_\a + c^2\, \nu^2\,,
\ee
where the left-invariant 1-forms of $SU(n+1)$ are defined in appendix
\ref{contractappx}.  The conditions for Ricci-flatness for the line bundle over
$\CP^n$ then follow from the Lagrangian $L=T-V$, where
\be
T = 2 \a'\, \gamma' + (2n-1)\, {\a'}^2\,,\qquad V= a^{4n-4}\, c^4
+ 2(n+1)\, a^{4n-2}\, c^2\,,
\ee
$d\eta= dt/(a^{2n}\, c)$, $a=e^\a$ and $b=e^{\beta}$.
The superpotential $W$ is given by
\be
W = a^{2n-2}\, c^2 + \fft{n+1}{n}\,a^{2n}\,.
\ee

   The scalings (\ref{cpnscal}) induce the following scalings
in the metric coefficients:
\be
a\longrightarrow \lambda^{-1}\, a\,,\qquad c\longrightarrow
\lambda^{-2}\, c\,.
\ee
After sending $\lambda$ to zero, the rescaled superpotential becomes
\be
W =  a^{2n-2}\, c^2\,.
\ee
Solving the resulting first-order equations gives
\be
a\propto t^{1/(n+1)}\,,\qquad c\propto t^{-n/(n+1)}\,.
\ee
In particular, for $n=1$ we recover the 4-dimensional Heisenberg metric
(\ref{4met}), and for $n=2$ we recover the 6-dimensional metric
(\ref{6met}). 

   It should be remarked that we could in fact obtain the same
Heisenberg contractions if the $\CP^n$ metrics are replaced by any
other $(2n)$-dimensional homogeneous Einstein-K\"ahler metrics of
positive scalar curvature.  In section \ref{6metricsec}, we shall give
a version of this construction for inhomogeneous Einstein-K\"ahler
manifolds.

\subsubsection{Contraction of $T^*S^{n+1}$}

   Starting from the left-invariant $SO(n+2)$ 1-forms in the notation
of (\ref{son2one}), the ansatz that gives rise to the Stenzel
\cite{stenzel} metrics on $T^*S^{n+1}$ is \cite{cglpsten}
\be
ds^2 = dt^2 + a^2\, \sigma_i^2 + b^2\, \td\sigma_i^2 + c^2\, \nu^2\,.
\ee
The Ricci-flat equations can be derived from the Lagrangian $L=T-V$
with
\bea
T &=& \a'\, \gamma' + \beta'\, \gamma' + n\, \a'\, \beta' +
\ft12(n-1)\, ({\a'}^2 + {\beta'}^2)\,,\nn\\
V&=& \ft14 (a\, b)^{2n-2}\, (a^4+b^4+c^4-2a^2\, b^2 - 2n\, (a^2+b^2)\,
c^2)\,,
\eea
and $V$ can be obtained from the superpotential \cite{cglpsten}
\be
W = \ft12 (a\, b)^{n-1}\, (a^2+b^2+c^2)\,.
\ee
Solutions of the associated first-order equations give the Ricci-flat
K\"ahler Stenzel metrics on $T^*S^{n+1}$ \cite{cglpsten}.

     After applying the scalings (\ref{son2scale2}), which imply
$(a,b,c) \longrightarrow (\lambda^{-2}\, a, \lambda^{-1}\, b,
\lambda^{-1}\, c)$,  and then sending
$\lambda$ to zero, we obtain the superpotential
\be
W = \ft12 a^{n+1}\, b^{n-1}\,.
\ee
This leads to the first-order equations \cite{cglpsten}
\be
\dot a = -\fft{a^2}{2b\, c}\,,\qquad \dot b= \fft{a}{2c}\,,\qquad 
\dot c = \fft{n\, a}{2b}\,,
\ee
and after defining a new radial variable by $dt= 2b\, c\, d\rho$, we
obtain the Ricci-flat Heisenberg metric
\be
ds^2 = \rho^{2n+2}\, d\rho^2 + \fft1{\rho}\, \sigma_i^2 + \rho\,
\td\sigma_i^2 + \rho^n\, \nu^2\,,
\ee
where the left-invariant 1-forms satisfy the exterior algebra
(\ref{son2alg2}).  Setting $n=2$, it is easily seen after a coordinate
transformation that we reproduce the Ricci-flat metric
(\ref{cy2met}).

\subsection{Contractions of 7-metrics of $G_2$ holonomy}

    In the present section, we shall show how the two
seven-dimensional Heisenberg metrics (\ref{7met}) and (\ref{7met2})
can be obtained as contraction limits of complete $G_2$ metrics of
cohomogeneity one.  Later, in section \ref{g2sec}, we shall give
a version of this construction using inhomogeneous hyper-K\"ahler
4-metrics.

\subsubsection{Contraction of $\R^3$ bundle over $S^4$}

   The complete metric of $G_2$ holonomy is \cite{brysal,gibpagpop}
\be
ds_7^2 = \Big(1+\fft{Q}{r^4}\Big)^{-1}\, dr^2 + r^2\, 
  \Big(1+\fft{Q}{r^4}\Big)\, (R_1^2+R_2^2)  + \ft1{2}\, r^2\, P_\a^2\,,
\label{g21bs}
\ee
where $R_1$, $R_2$ and $\sigma_\a$ are given in terms of the
left-invariant 1-forms of $SO(5)$ in appendix B.2.3.  After
implementing the rescalings given in (\ref{r1r2scal}), together with
\be
r\longrightarrow \lambda^{-2}\, r\,,\qquad 
Q\longrightarrow \lambda^{-12}\, Q\,,
\ee
then after sending $\lambda$ to zero we get the Heisenberg metric
\be
ds_7^2 =\fft{r^4}{Q}\, dr^2 +  
\fft{Q}{r^2}\, (\nu_1^2+\nu_2^2)  + \ft1{2}\, r^2\, \sigma_\a^2\,,
\label{g21bsh}
\ee
where $\nu_1$, $\nu_2$ and $\sigma_\a$ satisfy the contracted algebra
given in (\ref{r1r2alg}).  After a coordinate transformation, this can
be seen to be equivalent to the Heisenberg metric (\ref{7met}).

\subsubsection{Contraction of $\R^4$ bundle over $S^3$}

   The complete metric of $G_2$ holonomy is \cite{brysal,gibpagpop}
\be
ds_7^2 = \Big(1+\fft{Q}{r^3}\Big)^{-1}\, dr^2 + \ft19 r^2\, 
  \Big(1+\fft{Q}{r^3}\Big)\, \nu_i^2 + \ft1{12}\, r^2\, \sigma_i^2\,,
\label{g22bs}
\ee
in the notation of (A.2.2).  After taking the scaling limit
(\ref{so4scal}), together with
\be
r\longrightarrow \lambda^{-1}\, r\,,\qquad Q\longrightarrow
\lambda^{-5}\, Q\,,
\ee
and then sending $\lambda$ to zero, we obtain the following Heisenberg
limit of the metric (\ref{g22bs}):
\be
ds_7^2 = \fft{r^3}{Q}\, dr^2 + 
 \fft{Q}{9r}\, \nu_i^2 + \ft1{12}\, r^2\, \sigma_i^2\,,
\label{g22bsh}
\ee
where $\nu_i$ and $\sigma_i$ now satisfy the contracted exterior
algebra given in (\ref{g22heis}).  After a coordinate transformation,
this can be seen to be equivalent to the Heisenberg metric
(\ref{7met2}).

\subsection{Contraction of 8-metric of Spin(7) holonomy}

   Here, we shall show how the eight-dimensional Heisenberg metric
(\ref{d8metric}) can be obtained as a contraction limits of a complete
Spin(7) metric of cohomogeneity one.  Later, in section
\ref{8metricsec}, we shall give a version of this construction using
inhomogeneous hyper-K\"ahler 4-metrics.

   The complete metric of Spin(7) holonomy is \cite{brysal,gibpagpop}
\be
ds_8^2 = \Big(1+\fft{Q}{r^{10/3}}\Big)^{-1}\, dr^2 + 
\fft{9r^2}{100}\,  \Big(1+\fft{Q}{r^{10/3}}\Big)\, R_i^2 +
   \fft{9 r^2}{20}\, P_\a^2\,,\label{spin7bs}
\ee
where $L_i$ and $P_\a$ are given in terms of the left-invariant
1-forms of $SO(5)$ in appendix (A.2.3).  Implementing the scalings in 
(\ref{jkscal2}), together with
\be
r\longrightarrow \lambda^{-2}\, r\,,\qquad Q\longrightarrow
\lambda^{-32/3}\, Q\,,
\ee
then after sending $\lambda$ to zero the metric (\ref{spin7bs}) becomes
\be
ds_8^2 = \fft{r^{10/3}}{Q}\, dr^2 + 
\fft{9Q}{100 r^{4/3}}\, \nu_i^2 +
   \fft{9 r^2}{20}\, \sigma_\a^2\,,\label{spin7bsh}
\ee
where $\nu_i$ and $\sigma_\a$ now satisfy the contracted algebra 
(\ref{jkext2}).
After a coordinate transformation, this can be seen to be equivalent
to the $D=8$ Heisenberg metric (\ref{d8metric}).

\section{More general constructions of special-holonomy manifolds 
    in 6, 7 and 8 dimensions}
 
    It is clear from the structure of the Ricci-flat Heisenberg
metrics in dimensions 6, 7 and 8 in section \ref{dwsec} that in each
case where the principal orbits are torus bundles over $T^4$, this
4-torus can itself be replaced by an arbitrary Ricci-flat K\"ahler
4-metric.  In other words, we can allow the 4-manifold to be any
hyper-K\"ahler space.  Such a space admits a triplet of
covariantly-constant 2-forms $J^a$, which satisfy the multiplication
rules of the imaginary unit quaternions:
\be
J^a_{ij}\, J^b_{jk} = -\delta_{ab}\, \delta_{ij} + \ep_{abc}\,
J^c_{ik}\,.
\ee
In this section, we shall consider this more general construction in
each of the dimensions 6, 7 and 8.

\subsection{6-metric of $SU(3)$ holonomy from $T^1$ bundle over
hyper-K\"ahler\label{6metricsec}}

\subsubsection{Description in real coordinates}
 
   Let $ds_4^2$ be a hyper-K\"ahler 4-metric, and then consider the
following 6-metric:
\be
d\hat s_6^2 = H^2\, dy^2 + H^{-2}\, (dz_1 + A_\1)^2 + H\, ds_4^2\,,
\ee
where $H=y$, and $dA_\1=J$, a K\"ahler form on $ds_4^2$ (we take
$m=1$ here).  In the orthonormal frame
\be
\hat e^0 = H\, dy\,,\qquad \hat e^1 = H^{-1}\, (dz_1 + A_\1)\,,\qquad
\hat e^i =H^{1/2}\, e^i\,,
\ee
where $e^i$ is an orthonormal frame for $ds_4^2$, we find that the
spin connection is given by
\bea
&&\hat \omega_{01} =  H^{-2}\, \hat e^1\,,\qquad
\hat \omega_{0i}= -\ft12 H^{-2}\, \hat e^i\,,\qquad
\hat\omega_{1i} = \ft12 H^{-2}\, J_{ij}\, \hat e^j\,,\nn\\
&&\hat\omega_{ij} = \omega_{ij} - \ft12 H^{-2}\, J_{ij}\, \hat e^1\,,
\eea
where $\omega_{ij}$ is the spin connection for $ds_4^2$.
 
    From this, it follows that the curvature 2-forms are given by
\bea
\Theta_{01} &=& -3H^{-4}\, \hat e^0\wedge \hat e^1 + \ft32
H^{-4}\, J_{ij}\, \hat e^i\wedge \hat e^j \,,\nn\\
\Theta_{0i} &=& \ft34 H^{-4}\, \hat e^0\wedge \hat e^i + \ft34
H^{-4}\, J_{ij}\, \hat e^1\wedge \hat e^j \,,\nn\\
\hat \Theta_{1i} &=& \ft34 H^{-4}\, \hat e^1\wedge \hat e^i -
\ft34H^{-4}\, \hat e^0\wedge \hat e^j\,,\\
\hat\Theta_{ij} &=& \Theta_{ij} -\ft14 H^{-4}\, (\delta_{ik}\,
\delta_{j\ell} + J_{ik}\, J_{j\ell} + J_{ij}\, J_{k\ell})\, \hat e^k
\wedge \hat e^\ell\,,\nn
\eea
where $\Theta_{ij}$ is the curvature 2-form for $ds_4^2$.  From these,
we can read off that the Ricci tensor vanishes.
 
   The K\"ahler form for the 6-dimensional metric is given by
\be
\hat J = \hat e^0\wedge \hat e^1 + H\, J\,.
\ee
 
    The Lorentz-covariant exterior derivative $\hat D$ acting on a
spinor $\psi$ is given by
\bea
\hat D\, \psi &\equiv& d\psi + \ft14\, \hat\omega^{AB}\,
\hat\Gamma_{AB}\, \psi \,,\nn\\
&=& D\, \psi + \ft12 H^{-2}\, (\hat\Gamma_{01} - \ft14
J_{ij}\, \hat\Gamma_{ij})\, \psi\, \hat e^1 - \ft14H^{-2}\, (
\hat\Gamma_{0i} - J_{1j}\, \hat\Gamma_{1j})\, \psi\, \hat e^j\,,
\eea
where $D\equiv d + \ft14\omega^{ij}\, \hat\Gamma_{ij}$ is the
Lorentz-covariant exterior derivative on $ds_4^2$ (except that the
gamma matrices are the six-dimensional ones).
 
    It follows from this expression for $\hat D$ that a Killing spinor
$\hat \eta$ must satisfy the conditions
\be
D\, \hat\eta =0\,,\qquad  \hat\Gamma_{0i} \, \eta = J_{ij}\,
\hat\Gamma_{1j}\, \hat\eta\,.
\ee

\subsubsection{Description in complex coordinates}

   The above discussion made use of real coordinates on the
six-dimensional Ricci-flat K\"ahler manifold.  The structure of the
metric in the complex notation (\ref{6met4}), and of the K\"ahler
potential in the form (\ref{kah4}), suggest the natural generalisation
for the construction in a complex notation.  Thus we are led to the
following:

     Let $ds^2$ be a Ricci-flat K\"ahler metric of complex dimension
$n$, with K\"ahler function $K$, and K\"ahler form $J=\im\,
\del\bar\del\, K$.  Then
\be
d\td s^2 = H\, ds^2 + H^{-n}\, |d\zeta_{n+1} + A|^2\label{n1met}
\ee
is a Ricci-flat K\"ahler metric of complex dimension $(n+1)$, where
\be
A = \fft1{n+1}\, \del \, K   \,,\qquad H = \phi^{1/(n+1)}\,,
\ee
and we have defined
\be
\phi \equiv  1 + \zeta_{n+1} + \bar\zeta_{n+1} + \fft1{n+1}\, K\,.
\ee

(The ``1'' is inessential here, of course.)
The K\"ahler function for $d\td s^2$ is given by
\be
\wtd K = \fft{(n+1)^2}{n+2}\, \phi^{(n+2)/(n+1)} =
\fft{(n+1)^2}{n+2}\, H^{n+2}\,,
\ee
and its K\"ahler form is given by
\be
\wtd J = H\, J + \im\, H^{-n}\, (d\zeta_{n+1} + A) \wedge
(d\bar\zeta_{n+1} + \bar A)\,.
\ee
 
   The proof is as follows.  First, note that calculating $\wtd J$
from the K\"ahler function $\wtd K$ given above, we get
\bea
\wtd J &=&\im\, \del\bar\del\, \wtd K = \im\, \del\bar\del\, \Big(
\fft{(n+1)^2}{n+2}\, \phi^{(n+2)/(n+1)}\Big)\,,\nn\\
&=& \im\, (n+1)\, \del\, (\phi^{1/(n+1)}\, \bar\del\, \phi) =
\im\, (n+1)\, \phi^{1/(n+1)}\, \del\bar\del\, \phi + \im\,
\phi^{-n/(n+1)}\, \del\phi\wedge \bar\del\phi\,,\nn\\
&=& \im\, \phi^{1/(n+1)}\, \del\bar\del\, K + \im\, \phi^{-n/(n+1)}\,
(d\zeta_{n+1} + A)\wedge (d\bar \zeta_{n+1} + \bar A)\,,\nn\\
&=& H\, J + \im\, H^{-n}\, (d\zeta_{n+1} + A)\wedge (d\bar \zeta_{n+1}
+ \bar A)\,.
\eea
Bearing in mind that the K\"ahler form is related to the metric by
$\wtd J = \im\, \wtd g_{\mu\bar\nu}\, d\zeta^\mu\wedge \bar d\zeta^{\bar
\nu}$, we see that this does indeed agree with the metric given in
(\ref{n1met}).
 
  This shows that the metric (\ref{n1met}) is indeed K\"ahler.
Finally, to show that it is Ricci-flat, we calculate the determinant:
\be
\det(\td g) = H^{2n}\, H^{-2n}\,\det(g) = \det(g)\,.
\ee
Since the Ricci form is given by $\wtd {\cal R} = \im\, \del\bar\del\,
\log\det(\td g)$, it follows that if the Ricci form ${\cal R}$ for the
metric $ds^2$ is zero (which was the initial assumption), then the
Ricci form $\wtd {\cal R}$ for $d\td s^2$ is zero also.
 
   It is easily seen that the 6-metric metric we obtained in
section \ref{t1overt4ssec} is an example of this type, since the K\"ahler
function for the flat 4-torus can be taken to be $K=|\zeta_1|^2 +
|\zeta_2|^2$.

\subsection{7-metric of $G_2$ holonomy from $T^2$ bundle over 
hyper-K\"ahler}\label{g2sec}

    Let $ds_4^2$ be a hyper-K\"ahler metric, with a triplet of K\"ahler
forms $J^a$, with associated 1-form potentials $A_\1^a$:
\be
J^a = d A_\1^a\,,\qquad \nabla\, J^a=0\,.
\ee
It turns out to be convenient to let $a$ range over the values 0,1,2.
 
   Consider the metric
\be
d\hat s_7^2 = H^4\, dy^2 + H^{-2}\, \sum_{\a=1}^2 (dz^\a +
A_\1^\a)^2 + H^2\, ds_4^2\,,
\ee
where $H=y$.  (We have set $m=1$.)   Define vielbeins by
\be
\hat e^0=H^2\, dy\,,\qquad \hat e^\a = H^{-1}\, (dz^\a +
A_\1^\a)\,,\qquad \hat e^i = H\, e^i\,,
\ee
where $a=(0,\a)$, $i=(3,4,5,6)$, and $e^i$ is a vielbein for the
hyper-K\"ahler 4-metric $ds_4^2$.  Then it can be verified that the
following 3-form is closed:
\be
\psi_\3 \equiv \hat e^0\wedge \hat e^1\wedge \hat e^2 + H^2\, \hat
e^0\wedge J^0 -\ep_{\a\beta} \hat e^\a\wedge J^\beta\,.
\ee
 
   In fact this 3-form is covariantly constant, as can be verified
using the expressions for the spin connection:
\bea
&&\hat\omega_{0\a} = H^{-3}\, \hat e^\a\,,\qquad \hat \omega_{0i} =
-H^{-3}\, \hat e^i\,,\qquad \hat\omega_{\a\beta} =0\,,\nn\\
&&\hat\omega_{\a i} = \ft12 H^{-3}\, J^\a_{ij}\,
\hat e^i\,,\qquad
\hat\omega_{ij} = \omega_{ij} - \ft12 H^{-3}\,
J^\a_{ij}\, \hat e^\a\,,\label{d7spin}
\eea
where $J^a_{ij}$ denotes the components of $J^a$ with respect to the
vielbein $e^a$ for $ds_4^2$, and $\omega_{ij}$ is the spin-connection
for the vielbein $e^i$.  The covariant-constancy of $\psi_\3$ proves
that the metric $d\hat s_7^2$ has $G_2$ holonomy, and it is
also therefore Ricci flat.
 
    From (\ref{d7spin}) is is also straightforward to calculate the
vielbein components of the Riemann tensor for the metric $d\hat
s_7^2$.  We find
\bea
&&\hat R_{0\a 0\beta} = -4 H^{-6}\, \delta_{\a\beta}\,,\qquad
\hat R_{0\a ij} = 2H^{-6}\, J^\a_{ij}\,,\nn\\
&&\hat R_{0i0i} = 2H^{-6}\, \delta_{ij} \,,\qquad
\hat R_{0i\a j} =H^{-6}\, J^\a_{ij}\,,\nn\\
&&\hat R_{\a\beta\gamma\delta} = -H^{-6}\, (\delta_{\a\gamma}\,
\delta_{\beta\delta} - \delta_{\a\delta}\, \delta_{\beta\gamma})\,,
\qquad
\hat R_{\a\beta ij} = \ft12 H^{-6}\, \ep_{a\beta}\, J^0_{ij}\,,\nn\\
&& \hat R_{\a i \beta j} = \ft54 H^{-6}\, \delta_{\a\beta} \,
\delta_{ij} + \ft14 H^{-6}\, \ep_{\a\beta}\, J^0_{ij}\,,\\
&&\hat R_{ijk\ell} = H^{-2}\, R_{ijk\ell} - \ft14 H^{-6}\,
(J^\a_{ik}\, J^\a_{j\ell} - J^\a_{i\ell}\, J^\a_{jk} + 2 J^\a_{ij}\,
J^\a_{k\ell} + 4\delta_{ik}\, \delta_{j\ell} - 4\delta_{i\ell}\,
\delta_{jk})\,,\nn
\eea
where $R_{ijk\ell}$ is the Riemann tensor of the hyper-K\"ahler metric
$ds_4^2$.  It is easily verified from these expressions that the Ricci
tensor $\hat R_{AB}$ of the metric $d\hat s_7^2$ is zero.

\subsection{8-metric of Spin(7) holonomy from $T^3$ bundle over 
hyper-K\"ahler}\label{8metricsec}
 
    In a similar fashion, we can give the general construction for
8-metrics, in terms of a hyper-K\"ahler base metric $ds_4^2$.  This
time we shall have
\be
d\hat s_8^2 = H^6\, dy^2 + H^{-2}\, \sum_{a=1}^3 (dz^a + A_\1^a)^2 +
H^3\, ds_4^2\,.
\ee
Note that here, it is convenient to label the three K\"ahler forms of
$ds_4^2$ by $J^a=dA_\1^a$ with $a=1,2,3$.  We then define the
vielbeins
\be
\hat e^0 = H^3\, dy\,,\qquad \hat e^a = H^{-1}\, (dz^a +
A_\1^a)\,,\qquad \hat e^i = H^{3/2}\, e^i\,,
\ee
where here $i=4,5,6,7$, and $e^i$ is a vielbein for the
hyper-K\"ahler metric $ds_4^2$.  It can then be verified that the
self-dual 4-form $\Psi_\4$ given by
\be
\Psi_\4 = \hat e^0\wedge \hat e^1\wedge \hat
e^2\wedge \hat e^3 + \ft16 H^6\, J^a\wedge J^a + H^3(\hat e^0\wedge
\hat e^a + \ft12 \ep_{abc}\, \hat e^b \wedge \hat e^c)\wedge J^a
\ee
is closed.  (Note that $\ft16 J^a\wedge J^a$ is just another way of
writing the volume form of $ds_4^2$.)
 
   In fact it can also be verified that $\Psi_\4$ is covariantly constant,
by making use of the following results for the spin connection of the
8-metric:
\bea
&&\hat\omega_{0a}= H^{-4}\, \hat e^a\,,\qquad
  \hat\omega_{0i} = -\ft32 H^{-4}\, \hat e^i\,,\qquad
\hat\omega_{ab} =0\,,\nn\\
&&\hat\omega_{ai} = \ft12 H^{-4}\, J^a_{ij}\, \hat e^j\,,\qquad
\hat\omega_{ij} = \omega_{ij} -\ft12 H^{-4}\, J^a_{ij}\, \hat
e^a\,.
\eea
    Calculating the curvature from this, we find
\bea
&&\hat R_{0a0b} =-5H^{-8}\, \delta_{ab}\,,\qquad \hat R_{0aij} = \ft52
H^{-8}\, J^a_{ij}\,,\nn\\
&&\hat R_{0i0j} = \ft{15}{4}\, H^{-8}\, \delta_{ij}\,,\qquad
\hat R_{0iaj} = \ft54 H^{-8}\, J^a_{ij}\,,\nn\\
&&\hat R_{abcd} = -H^{-8}\, (\delta_{ac}\, \delta_{bd} - \delta_{ad}\,
\delta_{bc})\,,\qquad \hat R_{abij} = \ft12 H^{-8}\, \ep_{abc}\,
J^c_{ij}\,,\\
&&\hat R_{aibj} =\ft74 H^{-8}\, \delta_{ab}\, \delta_{ij} + \ft14
H^{-8}\, \ep_{abc}\, J^c_{ij}\,,\nn\\
&&\hat R_{ijk\ell} = H^{-2}\, R_{ijk\ell} -\ft14 H^{-8}\, (J^a_{ik}\,
J^a_{j\ell} -J^a_{i\ell}\, J^a_{jk} + 2J^a_{ij}\, J^a_{k\ell} +
9\delta_{ik}\, \delta_{j\ell} -9 \delta_{i\ell}\, \delta_{jk})\,.\nn
\eea
It is easily verified that the Ricci tensor $\hat R_{AB}$ for the
8-dimensional metric $d\hat s_8^2$ vanishes.

\section{Cosmological resolutions}

   There is an alternative approach to resolving the various
Heisenberg metrics that we have been discussing in this paper.  This
involves modifying the requirement of Ricci-flatness, so that instead
the metrics are now required to satisfy the Einstein condition with a
negative Ricci tensor.  It turns out in all the previous examples, we
can now obtain complete and non-singular non-compact metrics.  In each
case, this is achieved by replacing the various powers of $H$
appearing as prefactors of the terms $(dz_i+\cdots)^2$ by arbitrary
functions of the radial variable, and then solving the Einstein
equations.

    In all the cases we consider, the homothetic conformal Killing
vector $D$ of the original Ricci-flat metric is replaced by a true Killing
vector of the associated Einstein metric.  This, together with the
generators of the nilpotent Heisenberg group generate a solvable
group, which acts simply-transitively on the Einstein manifold, which
may thus be taken to be a solvable group manifold ${\rm Solv}$.  In
addition, all our metrics admit some manifest compact symmetries,
which act linearly on the Heisenberg manifold.  We have also
identified some non-linearly acting symmetries.  In all cases, we can
express the manifold as $G/H={\rm Solv}$, where the non-compact group
$G$ has maximal compact subgroup $H$.  The group $H$ contains the
linearly-realised compact symmetries.  This is quite striking because
a theorem of Alekseevskii and Kimel'fel'd} \cite{alekim} states that
any homogeneous non-compact Ricci flat Riemannian metric must be flat
\cite{heber}. In fact a theorem of Dotti states that a left-invariant
Einstein metric on a unimodular solvable group must be flat, so our
solvable groups cannot be unimodular, that is the trace of the
structure constants of the Lie algebra cannot vanish
\cite{dotti}.\footnote{We shall demonstrate this explicitly below for
all our examples.}

   The simplest example is when the Ricci-flat manifold is flat space,
and the associated solvable group manifold is hyperbolic space.  This
has been encountered in studies of the AdS/CFT correspondence and is
related to the ideas of Ref. \cite{guklpo}, in which the fifth
dimension corresponds to the Liouville mode of a non-critical string
theory which thus becomes dynamical.  The idea is that the string
coordinates appear in the effective action multiplied by a function of
the Liouville field. This function should vanish at large negative
values of the Liouville field, in order to enforce a ``zig-zag
symmetry.'' To achieve this and to fix the functional form, the
effective Lagrangian for the string is taken to include a piece
invariant under both Poincar\'e transformations and dilatations.  As a
result of imposing the dilatation symmetry, an exponential function of
the Liouville mode multiplies the coordinates of the string. The
vanishing of this function corresponds to the horizon in AdS
spacetime.  From the ten-dimensional point of view, one must take the
product metric AdS$_5\times S^5$, where the $SO(6)$ R-symmetry group
arises from the isometry group of the $S^5$ factor.

   Our metrics arise by replacing the usual commuting translations of
the string by non-commuting translations satisfying a Heisenberg
algebra.  This may be relevant when considering strings in constant
background fields.  It is a striking fact that the obvious nilpotent
symmetry is, as in the standard AdS case, enhanced to a much larger
group $G$.
 
   A common feature of all of our Ricci-flat metrics is that the size
of the toric fibres goes to zero as a negative power of distance as we
go to infinity, while the size of the base expands as a positive
power.  By contrast, for our Einstein metrics both directions expand
exponentially as one goes to infinity, but the toric fibre directions
expand more rapidly than the base.  In some cases the exponential
expansion of some of the directions in the base is different from that
of other directions.  This has the consequence that the conformal
geometry on the boundary is singular. If one were to use as conformal
factor the scale-size of the fibres, then the metric on the base would
tend to zero. If one used the metric on the smallest-growing base
direction, then the metric on the fibres would diverge. In the case of
a single scale-factor, the resulting metric is referred to by
mathematicians as a Carnot-Carath\'eodory metric \cite{biq}.  This
behaviour has been commented on in Ref. \cite{stronger} and
\cite{tayrob} in the case of the four-dimensional Bergman metric on
$SU(2,1)/(SU(2)\times U(1))$. In these references the metric was
written in coordinates adapted to the maximal compact subgroup. At
constant radius the metric is a squashed 3-sphere, where the ratio of
lengths on the $U(1)$ fibres compared with the $S^2$ base diverges as
one approaches infinity. In fact, as we shall illustrate below, one
may also write the metric in Heisenberg-horospherical coordinates and
obtain the same behaviour. In other words, the Bergman metric is of
cohomogeneity one with respect to {\sl both} $SU(2)$ and its
contraction to the $\frak{ur}$-Heisenberg group.

    In order to get a solution of Type IIB theory in ten dimensions,
one needs a five-dimensional rather than a four-dimensional metric.
In what follows we shall present a new complete five-dimensional
Einstein metric on a solvable group manifold, which may be used to
obtain a Euclidean-signatured solution of Type IIB supergravity in ten
dimensions, with a complex self-dual 5-form. (This theory is obtained
by Wick rotation from the Lorentzian IIB theory, and the components of
the 5-form with a time index are purely imaginary.)  In general, these
solutions need have no real Lorentzian sections.  Although the
five-dimensional metric may well have appeared before in general
mathematical classification schemes, it and all of our metrics that
are not symmetric spaces have not, as far as we are aware, been
previously written down explicitly, nor have they been used in
the construction of supergravity solutions.
 
   Physically, the unusual behaviour of the boundary appears to be
related to a mismatch in dimension between the boundary theory and the
dimension of the bulk theory minus one.  This behaviour presumably
arises because, from the Kaluza-Klein point of view, a Heisenberg
isometry gives rise to a background magnetic field.  Systems in strong
magnetic fields are well known to exhibit a reduction in
dimensionality.  The size of the toric fibre here is inversely
proportional to the electric charge, and so as we go to infinity in
our metrics the charge goes to zero, at constant magnetic field.
Equivalently, the magnetic field goes to infinity, at fixed electric
charge.
   
   In some instances, we can give a more complete interpolation
between a Ricci-flat Heisenberg metric and the associated
``cosmological resolution;'' these are constructed in section
\ref{heidesec}.  Specifically, for the four-dimensional metric
(\ref{4met}), and for the two six-dimensional metrics (\ref{6met}) and
(\ref{cy2met}), we can obtain more general solutions with both a
charge parameter and a cosmological constant.  Indeed, in the case of
(\ref{4met}) and (\ref{6met}) these ``Heisenberg-de Sitter'' metrics
are themselves specialisations of already known metrics with
cosmological constants.  Thus the four-dimensional Heisenberg-de Sitter
metric is a contraction limit of the Eguchi-Hanson-de Sitter metric
\cite{gibpop,ehds}, and the six-dimensional Heisenberg-de Sitter
generalisation of (\ref{6met}) is a contraction of a complete metric
with cosmological constant on the complex line bundle over $\CP^2$.
General results for such cosmological metrics on line bundles over
Einstein-K\"ahler spaces were obtained in \cite{pagpop1}.  We expect
that the Heisenberg-de Sitter generalisation of the six-dimensional
metric (\ref{cy2met}) that we obtain in section \ref{heidesec} may
similarly be a contraction limit of a ``Stenzel-de Sitter'' metric.

\subsection{The Einstein metrics}

   We shall begin by listing the results Einstein metrics for all the
cases. As remarked above, in the original Ricci-flat metrics the
lengths of the Kaluza-Klein fibre directions go to zero at large $y$
while the base space expands.  By contrast, in the related Einstein
metrics both the fibre and base-space directions expand exponentially
as one goes to infinity.  In fact the fibre directions now expand
faster than the base.  After each metric, we give its Ricci tensor,
and also the algebra of exterior derivatives of the vielbein 1-forms.
We choose the obvious basis, with $ds^2= e^a\otimes e^a$, and
$e^0=dt$, {\it etc}.
\medskip

\noindent\underline{$D=4$; $T^1$ bundle over $T^2$:}

\bea
ds_4^2 &=& dt^2 + 4k^2\, e^{4 k\, t}\, (dz_1 + z_3\, dz_2)^2 + e^{2k\,
t}\, (dz_2^2 + dz_3^2)\,,\label{d4cos}\\
R_{ab} &=& -6k^2\, g_{ab}\,,\nn\\
de^0&=&0\,,\quad de^1=2k\, (e^0\wedge e^1-e^2\wedge e^3)\,,\quad 
de^2= k\, e^0\wedge e^2\,,\quad de^3=k\, e^0\wedge e^3\,.\nn
\eea
 
\bigskip

\noindent\underline{$D=5$; $T^1$ bundle over $T^3$:}

\bea
ds_4^2 &=& dt^2 + 22k^2\, e^{8k\, t}\, (dz_1 + z_3\, dz_2)^2 + e^{4k\,
t}\, (dz_2^2 + dz_3^2) + e^{6k\, t}\, dz_4^2\,,\label{d5cos}\\
R_{ab} &=& -  22 k^2\, g_{ab}\,,\nn\\
de^0&=& 0\,,\quad de^1=4k\, (e^0\wedge e^1 -e^2\wedge e^3) \,,\quad
de^2=2k\, e^0\wedge e^2\,,\quad de^3=2k\, e^0\wedge e^2\,,\nn\\
de^4&=& 3k\, e^0\wedge e^4\,.\nn
\eea

\bigskip

\noindent\underline{$D=6$; $T^1$ bundle over $T^4$:}

\bea ds_6^2 &=& dt^2 + 4k^2\, e^{4 k\, t}\, (dz_1 + z_3\, dz_2+ z_5\,
dz_4)^2 + e^{2k\, t}\, (dz_2^2 + dz_3^2+dz_4^2 + dz_5^2)\,,\label{d6cos1}\\
R_{ab} &=& -8k^2\, g_{ab}\,,\nn\\
de^0&=&0\,,\quad de^1=2k\, (e^0\wedge e^1 -e^2\wedge e^3-e^4\wedge
e^5) \,,\nn\\
de^2&=& k\, e^0\wedge e^2\,,\quad de^3= k\, e^0\wedge e^3\,,\quad 
de^4= k\, e^0\wedge e^4\,,\quad de^5= k\, e^0\wedge e^5\,.\nn 
\eea
 
\bigskip

\noindent\underline{$D=6$; $T^2$ bundle over $T^3$:}

\bea ds_6^2 &=& dt^2 + 36k^2\, e^{10k\, t}\,[ (dz_1 + z_4\, dz_3)^2 
+ (dz_2 + z_5\, dz_3)^2 ]\nn\\
&&
+ e^{4k\, t}\, dz_3^2 + e^{6k\, t}\, (dz_4^2 + dz_5^2)\,,\label{d6cos2}\\
R_{ab} &=& -18k^2\, g_{ab}\,,\nn\\
de^0&=&0\,,\quad de^1=5k\, (e^0\wedge e^1 -e^3\wedge e^4) \,,\quad
de^2= 5k\, (e^0\wedge e^2-e^3\wedge e^5)  
\,,\nn\\
 de^3&=& k\, e^0\wedge e^3\,,\quad 
de^4= k\, e^0\wedge e^4\,,\quad de^5= k\, e^0\wedge e^5\,. \nn
\eea
 
\bigskip

\noindent\underline{$D=7$; $T^2$ bundle over $T^4$:}

\bea ds_7^2 &=& dt^2 + 4k^2\, e^{4 k\, t}\, [(dz_1 + z_4\, dz_3+ z_6\,
dz_5)^2 +  (dz_2 + z_5\, dz_3- z_6\,
dz_4)^2]\nn\\
&& + e^{2k\, t}\, (dz_3^2 + dz_4^2+dz_5^2 + dz_6^2)\,,\label{d7cos1}\\
R_{ab} &=& -12k^2\, g_{ab}\,,\nn\\
de^0&=&0\,,\quad de^1=2k\, (e^0\wedge e^1 -e^3\wedge e^4-e^5\wedge
e^6) \,,\nn\\
de^2 &=& 2k\, (e^0\wedge e^2 -e^3\wedge e^5-e^6\wedge e^4 )
\,,\quad de^3= k\, e^0\wedge e^3\,,\quad 
de^4= k\, e^0\wedge e^4\,,\nn\\
 de^5&=& k\, e^0\wedge e^5\,,\quad de^6=k\, e^0\wedge e^6\,. \nn
\eea
 
\bigskip

\noindent\underline{$D=7$; $T^3$ bundle over $T^3$:}

\bea 
ds_7^2 &=& dt^2 + 6k^2\, e^{4 k\, t}\,[ (dz_1 + z_6\, dz_5)^2 
+ (dz_2 + z_6\, dz_4)^2 + (dz_3 + z_5\, dz_4)^2] \nn\\
&&+ e^{2k\, t}\, (dz_4^2 + dz_5^2+dz_6^2)\,,\label{d7cos2}\\
R_{ab} &=& -15k^2\, g_{ab}\,,\nn\\
de^0&=&0\,,\quad de^1=2k\, (e^0\wedge e^1 -e^5\wedge e^6) \,,\quad
de^2= 2k\, (e^0\wedge e^2-e^4\wedge e^6)  
\,,\nn\\
de^3 &=& 2k\, (e^0\wedge e^3 -e^4\wedge e^5)\,,\quad  
de^4= k\, e^0\wedge e^4\,,\nn\\
 de^5&=& k\, e^0\wedge e^5\,,\quad
de^6=k\, e^0\wedge e^6\,. \nn
\eea
 
\bigskip

\noindent\underline{$D=8$; $T^3$ bundle over $T^4$:}

\bea ds_8^2 &=& dt^2 + 4k^2\, e^{4 k\, t}\, [(dz_1 + z_5\, dz_4+ z_7\,
dz_6)^2 +  (dz_2 + z_6\, dz_4- z_7\,
dz_5)^2\nn\\
&&\qquad\qquad  + (dz_3+ z_7\, dz_4 + z_6\, dz_5)^2]\nn\\
&& + e^{2k\, t}\, (dz_4^2 + dz_5^2+dz_6^2 + dz_7^2)\,,\label{d8cos}\\
R_{ab} &=& -16k^2\, g_{ab}\,,\nn\\
de^0&=&0\,,\quad de^1=2k\, (e^0\wedge e^1 -e^4\wedge e^5-e^6\wedge
e^7) \,,\nn\\
de^2 &=& 2k\, (e^0\wedge e^2 -e^4\wedge e^6 +e^5\wedge e^7 )
\,,\quad de^3 =2k\, (e^0\wedge e^3 -e^4\wedge e^7 -e^5\wedge e^6
)\,,\nn\\
de^4&=& k\, e^0\wedge e^4\,,\quad
 de^5= k\, e^0\wedge e^5\,,\quad de^6=k\, e^0\wedge e^6\,,\quad
de^7=k\, e^0\wedge e^7\,.\nn
\eea

   Note that the algebras of exterior derivatives are all of the form 
$de^a = -\ft12 c^a{}_{bc}\,e^b\wedge e^c$, where the $c^a{}_{bc}$ are
constants.  These are in fact the structure constants of the
corresponding solvable groups.  Observe that these are indeed not
traceless, $c^a{}_{ba}\ne 0$, as is required by Dotti's theorem.  

   In the next subsection, we shall discuss these solvable groups as
coset spaces, exploiting the Iwasawa decomposition.

\subsection{Coset constructions}

\subsubsection{$SU(n,1)/U(n)=\wtd{\CP^n}={\rm H}_{\C}^n$}

    Those examples above whose principal orbits are of the form of
$T^1$ bundles over $T^p$ are in fact Bergman metrics on the
non-compact forms of $\wtd{\CP^n}$, with $p=2n-2$.  These are nothing
but the Fubini-Study metrics with the opposite sign for the
cosmological constant.  They are obtained by starting from coordinates 
$Z^A$ on $\C^{n+1}$, with the constraint
\be
\eta_{AB}\, Z^A\, \bar Z^B = -1\,,\label{ads}
\ee
where $\eta_{AB}$ is diagonal with $\eta_{00}= -1$, $\eta_{ab}=1$,
where $1\le a\le n$.   The Hopf fibration of this AdS$_{2n+1}$ by
$U(1)$ (taken to be timelike) then gives the Bergman metric.

    We can express the Bergman metric in a ``horospherical'' form
\cite{posasc},  
by introducing real coordinates $(\tau, \phi, \chi, x_i, y_i)$, in
terms of which we parametrize the $Z^A$ that satisfy (\ref{ads}) as
\bea
Z^0 &=& e^{\fft{\im}{2}\tau}\, \Big( \cosh\ft12\phi + 
\ft18 e^{\fft12\phi}\, (4\im\, \chi + x_i^2 + y_i^2)\Big)\,,\nn\\
Z^n &=& e^{\fft{\im}{2}\tau}\, \Big( \sinh\ft12\phi - 
\ft18 e^{\fft12\phi}\, (4\im\, \chi + x_i^2 + y_i^2)\Big)\,,\nn\\
Z^i &=& \ft12 e^{\fft{\im}{2}\tau + \fft12\phi}\, (x_i+\im\, y_i)\,,
\label{horo1}
\eea
where $1\le i\le n-1$.  Substituting into the metric $d\hat s^2
=\eta_{AB}\, dZ^A\, d\bar Z^B$ on AdS$_{2n+1}$, we find
\be
d\hat s^2= -\ft14\Big(d\tau + e^\phi\, [d\chi+\ft12(y_i\, dx_i
- x_i\, dy_i)]\Big)^2 + d\Sigma_{2n}^2\,,
\ee
where
\be
d\Sigma_{2n}^2 = \ft14 d\phi^2 + \ft14 e^\phi\, (dx_i^2+dy_i^2) 
         + \ft14 e^{2\phi}\,  [d\chi+\ft12(y_i\, dx_i
- x_i\, dy_i)]^2\,.\label{bergman}
\ee
Thus if we fibre AdS$_{2n+1}$ by the $U(1)$ whose coordinate $\tau$ is
the time parameter, we obtain the Bergman metric (\ref{bergman}) on
$\wtd{\CP^n}$. Comparing with (\ref{d4cos}) and (\ref{d6cos1}), we see
that these correspond to $\wtd{\CP^2}$ and $\wtd{\CP^3}$ respectively.

    If $n=3$, the denominator group $U(3)$ contains the
linearly-realised $U(2)$ noted in section \ref{t1overt4ssec}, 
and similarly if $n=2$ the denominator group $U(2)$ contains the
linearly-realised $U(1)$ noted in section \ref{egsec}.

   In the case $n=2$ one can regard this solution as a special case of
the Taub-NUT-de Sitter metrics, which have been applied to the AdS/CFT
correspondence in \cite{chemjomy}. (For more general higher-dimensional
metrics of this type, see \cite{awacha}.)  The case $n=2$ is of further
interest because, while the Bergman metric is not conformal to the
associated Ricci-flat metric\footnote{It is impossible for two Riemannian
Einstein metrics to be conformal with non-constant conformal factor.},
there is a metric which is conformal to our Ricci-flat metric that is
distinguished by the property that it is essentially the only
non-trivial complete homogeneous hyper-Hermitean metric \cite{barber}.
The conformally related metric is
\be
ds ^2 + z^{-4} (d\tau + xdy)^2 + z^{-2} (dx^2 + dy^2 + dz^2)
\ee
It would be interesting to know whether our other Ricci-flat
metrics are conformal to similarly-distinguished metrics.

\subsubsection{$Sp(n,1)/(Sp(n)\cdot Sp(1))= \wtd{\HP^n}={\rm H}_{\H}^n$}

    A similar construction can be given for the non-compact versions
of the quaternionic projective spaces, $\wtd{\HP^n}$.   Now, we start
from $n+1$ quaternionic coordinates $Q^A$, subject to the constraint
\be
\eta_{AB}\, Q^A\, \bar Q^B=-1\,,\label{ads3t}
\ee
where again $\eta_{AB}$ is diagonal with $\eta_{00}= -1$, $\eta_{ab}=1$,
where $1\le a\le n$.  This restricts us to a spacetime of anti-de
Sitter type, except that now we have three timelike coordinates.

   We can again introduce real horospherical coordinates.  The three
times appear as the Euler angles of $SU(2)$, and in fact we can just
introduce them implicitly via the $Sp(1)=SU(2)$ quaternionic $U$.  In
addition, we introduce real coordinates $(\phi,\chi_\a,x_i,y_i^\a)$,
where $1\le\a\le3$ and $1\le i\le n-1$.  We shall denote the
imaginary unit quaternions by $\iota_\a=(\im,\jm,\km)$ (in terms of
which $U$ can be written as $U=e^{\fft{\km}{2}\, t_1}\,
e^{\fft{\im}{2}\, t_2}\, e^{\fft{\km}{2}\, t_3}$).  We then parametrize
the quaternions $Q^A$ that satisfy (\ref{ads3t}) as 
\bea
Q^0 &=& U\, \Big( \cosh\ft12\phi + 
\ft18 e^{\fft12\phi}\, (4\iota_\a\, \chi_\a + x_i^2 + 
(y^\a_i)^2)\Big)\,,\nn\\
Q^n &=& U\, \Big( \sinh\ft12\phi - 
\ft18 e^{\fft12\phi}\, (4\iota_\a \,\chi_\a + x_i^2 + 
(y^\a_i)^2)\Big)\,,\nn\\
Q^i &=& \ft12 U\, e^{\fft12\phi}\, (x_i+\iota_\a\, y^\a_i)\,.
\label{horo2}
\eea
These are closely analogous to (\ref{horo1}) for the complex case.
Substituting into the metric $d\hat s^2 = \eta_{AB}\, dQ^A\, d\bar
Q^B$ on the three-timing AdS$_{4n+3}$, we find
\be
d\hat s^2 = -\Big| U^{-1}\, dU +  e^\phi\, \iota_\a\, 
        [d\chi_\a + \ft12(x_i\, dy^\a_i -y^\a_i\,  dx_i)]\Big|^2 + 
    d\Xi_{4n}^2\,,
\ee
where
\be
d\Xi_{4n}^2 = \ft14 d\phi^2 + \ft14 e^\phi\, (dx_i^2 + (dy_i^\a)^2) 
    +\ft1{4}\, e^{2\phi}\, [d\chi_\a + \ft12(x_i\, dy^\a_i -y^\a_i\,
dx_i)]^2\,.\label{bergmanh}
\ee
Thus if we project orthogonally to the $SU(2)$ timelike fibres, we
obtain (\ref{bergmanh}) as the metric on $\wtd{\HP^n}$.  It is the
coset $Sp(n,1)/(Sp(n)\cdot Sp(1))$.

   Comparing with (\ref{d8cos}), we see that our 8-dimensional
Einstein metric is precisely the non-compact ``quaternionic Bergman
metric'' on $\wtd{\HP^2}$, which is the coset $Sp(2,1)/(Sp(2)\cdot
Sp(1))$.  In this case the denominator group contains the
linearly-realised $Sp(1)\equiv SU(2)$ noted in section \ref{t1overt3ssec}.  

   In all cases, one may check that the exponential expansion of the
$SU(2)$ fibres is more rapid than that of the $\wtd{\HP^n}$ base, as
one goes to infinity.

\subsection{Heisenberg-de Sitter metrics}\label{heidesec}

   So far, we have considered Ricci-flat Heisenberg metrics, and also
``cosmological resolutions'' that do not have an immediate
mathematical relation to the Heisenberg metrics.  In certain cases, at
least, we can find a more general solution that encompasses both the
Heisenberg metric and the cosmological metric, as certain limits.  In
fact, at least in some of these examples, we know that there exist
``de Sitterised'' versions of the complete non-singular Ricci-flat
metrics, even before the Heisenberg limit is taken.

   A case in point is the Eguchi-Hanson-de Sitter metric, given by
\be
ds_4^2 = F^{-1}\, dt^2 + \ft14 r^2\, F\, \sigma_3^2 + \ft14 r^2\,
(\sigma_1^2 + \sigma_2^2)\,,
\ee
where
\be
F = 1 + \fft{Q}{r^4} - \ft16\Lambda\, r^2\,.
\ee
It is an Einstein metric, with $R_{ab}=\Lambda\, g_{ab}$, and so
$\Lambda$ is the cosmological constant.  

   If we now take the Heisenberg limit, as in section \ref{egheissec}, we
obtain the Heisenberg-de Sitter metric
\be
ds_4^2 = F^{-1}\, dt^2 + \ft14 r^2\, F\, (dz_1+ z_3\, dz_2)^2 + \ft14 r^2\,
(dz_2^2 + dz_3^2)\,,\label{4heisde}
\ee
where we are dropping the tildes used to denote the rescaled
quantities in section \ref{egheissec}, and now we have
\be
F = \fft{Q}{r^4} - \ft16\Lambda\, r^2\,.\label{4F}
\ee
Note that $\Lambda$ does not suffer any rescaling in the taking of
this limit, so the cosmological constant is still $\Lambda$.  

   Having obtained the Heisenberg de-Sitter 4-metric, we can note that
if we set $\Lambda$ to zero, we recover (after an obvious coordinate
transformation) the Heisenberg metric (\ref{4met}).  On the other
hand, if we set $Q$ to zero, then after an obvious coordinate
transformation, (\ref{4heisde}) gives the cosmological resolution
(\ref{d4cos}).  If we keep both $Q$ and $\Lambda$ non-vanishing, we
have a more general Einstein metric that encompasses both the
Heisenberg and cosmological metrics discussed previously.

   We can now attempt a generalisation of the above to other cases.
Analogues of the Eguchi-Hanson-de Sitter metric are known for all
Ricci-flat metrics on complex line bundles over Einstein-K\"ahler
bases \cite{pagpop1}.  Thus we can expect to be able to get
generalised Heisenberg-de Sitter metrics for all the cases where the
principal orbits are $T^1$ bundles over $T^{2n}$.  For example, in
$D=6$ we can get a Heisenberg-de Sitter metric for the case of $T^1$
bundle over $T^4$, namely
\be
ds_6^2= F^{-1}\, dr^2 + r^2\, F\, [dz_1 + m\, (z_3\, dz_2 +
z_5\, dz_4)]^2  + \ft12 r^2 \,  (dz_2^2 + \cdots +
 dz_5^2)\,,\label{6heisde1}
\ee
where
\be 
F = \fft{Q}{r^6} -\ft18\Lambda\, r^2\,.
\ee
This is equivalent to (\ref{6met}) if $\Lambda$ is set to zero.  On
the other hand, if $Q$ is instead set to zero, it is equivalent to the
cosmological resolution metric (\ref{d6cos1}).

   A slightly more complicated example is the second of the two $D=6$
Heisenberg metrics, given in (\ref{cy2met}).  Here, we find that the
following is an Einstein metric, with cosmological constant $\Lambda$:
\bea
ds_6^2 &=& F^{-1}\, dr^2  + r^2\, F^{2/3}\,[ 
(dz_1 + m\, z_4\, dz_3)^2 +
(dz_2 + m\, z_5\, dz_3)^2] \nn\\
&&+ r^2\,F^{-1/3}\,  dz_3^2 + \ft12 r^2\, (dz_4^2 + dz_5^2)
\,,\label{6heisde2}
\eea
where
\be
F = \fft{Q}{r^6} -\ft18 \Lambda\, r^2\,.
\ee
This again has the appropriate limits, yielding (\ref{cy2met}) if
$\Lambda$ is set to zero, and yielding (\ref{d6cos2}) if instead $Q$
is set to zero.

    Having obtained these Heisenberg-de Sitter metrics, we can observe
that they provide a way to ``cap off'' the large-radius portions of
the Ricci-flat Heisenberg metrics that are transverse to the
domain-wall spacetimes.  In this respect, they appear to be conjugate
to the constructions of \cite{gibryc}, which by contrast resolve the
curvature singularities in the small-radius portions of the
domain-wall spacetimes.

   Let us illustrate this by considering the example of the 4-metric
(\ref{4met}), transverse to the domain wall in $D=8$ supergravity.  If
we take the Heisenberg-de Sitter metric (\ref{4heisde}), with both $Q$
and the cosmological constant $\Lambda$ in (\ref{4F}) positive, we see
that the metric running from the singularity at $r=0$ reaches a
natural endpoint at $r=r_0$, where the function $F$ vanishes, \ie at
\be
r_0^6 = \fft{6Q}{\Lambda}\,.
\ee
To study the behaviour of the metric near $r=r_0$, we introduce a new
coordinate $\rho$ defined by $r=r_0-\rho^2$.  In terms of this, the
metric near $r=r_0$ takes the form
\be
ds_4^2 \sim \fft{2 r_0^5}{3Q}\, \Big( d\rho^2 + \fft{r_0^2}{16}\, 
\Big(\fft{6Q}{r_0^5}\Big)\, \rho^2\, (dz_1 + z_3\, dz_2)^2 \Big) +
\ft14 r_0^2 \, (dz_2^2+dz_3^2)\,.
\ee
This will be regular at $r=r_0$ provided that $z_1$ has a period given by
\be
\Delta\, z_1 = \fft{4\pi\, r_0^4}{3Q}\,.
\ee

\section{Conclusions}

   As mentioned in the introduction, one of the motivations for the
present study was the possibility of resolving some of the BPS domain
walls along the lines of the``single-sided" domain-wall construction
described in \cite{gibryc}.  The case studied in detail in that
reference was four-dimensional, and the resolution was a certain
non-compact degeneration of a K3 surface, but generalisations to
higher dimensions were also indicated there which would correspond,
for example, to non-compact degenerations of Calabi-Yau complex
3-folds and 4-folds.  For example, the metric based on a circle bundle
over a K3 surface seems to be related to the metric one would obtain
by solving the Monge-Amp\'ere equation on the complement of a quartic
surface in $\CP^3$. In the case of Calabi-Yau metrics, one has various
proofs showing the existence of smooth resolved metrics, but these
give very little detailed information about the explicit forms of the
metrics.  The information one gets is mainly about the asymptotic form
of the metric near infinity. This is where our work may help identify
the resolutions.  In the case of K3 surfaces, as well as
asymptotically ``nil-manifolds'' based on generalised Heisenberg
groups, one also encounters asymptotically ``solv-manifolds,'' based
on solvable groups.  It may be that our work in this paper and
generalisations of it may be relevant to higher-dimensional Calabi-Yau
spaces.  A more challenging problem would be to relate our metrics to
compact Calabi-Yau manifolds.  For reasons explained in the
introduction, it is not easy to see how to do this using cohomogeneity
one Ricci-flat metrics.  One may, of course, give qualitative
discussions \cite{aspinwall} but it is extremely hard to make
quantitative progress with present-day techniques.
 
    In the case of metrics with exceptional holonomy the situation is
much less clear than in the case of Calabi-Yau metrics, because
existence theorems have been studied to a much lesser extent. However,
the important work of Joyce described in his recent book
\cite{joycebook} encourages us to believe that our results will prove
applicable in that case as well.

\section*{Acknowledgements}

   We should like to thank Massimo Bianchi, Anna Ceresole, Mirjam
Cveti\v{c}, Isabel Dotti, Andre Lukas, Krystof Pilch, Toine van Proeyen, Simon
Salamon and Paul Tod for helpful conversations and information. In
particular we should like to thank Paul Tod for an essential remark
about conformally-related metrics.  Subsets of the authors would like
to thank Texas A\&M Physics Department (H.L., K.S.S.), Imperial
College (C.N.P.), CERN (G.W.G., H.L., C.N.P., K.S.S.), SISSA (C.N.P.,
K.S.S.), the Institut Henri Poincar\'e (G.W.G., C.N.P., K.S.S.),
 Michigan University Physics Department (C.N.P.), DAMTP
(C.N.P.), the Benasque Centre for Science (G.W.G., C.N.P., K.S.S.) and
the Ecole Normale (C.N.P., K.S.S.) for hospitality at various stages
during this work.

\appendix

\section{The Raychaudhuri equation}
 
Let $T^\alpha$ be a unit vector tangent to a congruence of curves,
\be
T^\alpha T_\alpha=1 \Rightarrow T_{\alpha;\beta}\, T^\alpha=0\ .
\ee
Let us decompose the covariant derivative of $T_\alpha$ perpendicular
and parallel to $T_\a$ using the projection operator
$h_{\alpha\beta}=g_{\alpha\beta}-T_\alpha T_\beta$, such that
$h_{\alpha\beta}T^\beta=0.$ One has the decomposition
\be
T_{\alpha;\beta}=\Theta_{\alpha\beta}
  +\omega_{\alpha\beta}+T_\beta \, a_\alpha\ ,
\ee
where $\Theta_{\alpha\beta}=\Theta_{\beta\alpha}$ is a symmetric
expansion tensor and $\omega_{\alpha\beta}=-\omega_{\beta\alpha}$ is
an antisymmetric vorticity tensor, both of which are orthogonal to
$T_\alpha$,
\be
\Theta_{\alpha\beta}T^\beta=\omega_{\alpha\beta}T^\beta=0.
\ee
The vector $a_\alpha=T_{\alpha;\beta}\, T^\beta$ is the {\em
acceleration} vector, and would vanish if the congruence is geodesic.

    Since $T_\alpha \, a^\alpha=0$, we have
\be
T^\alpha_{;\alpha}=\Theta_{\alpha\beta}\, h^{\alpha\beta}=
\Theta_{\alpha\beta}\, g^{\alpha\beta}.
\ee
This is the {\em expansion}. We may then set
\be
\Theta_{\alpha\beta}={\Theta h_{\alpha\beta}\over d} + 
\Sigma_{\alpha\beta}\ ,
\ee
where
$\Sigma_{\alpha\beta}$ is the {\em shear}, satisfying
$\Sigma_{\alpha\beta}h^{\alpha\beta}=
\Sigma_{\alpha\beta}g^{\alpha\beta}=0$.
Thus we have
\be
T_{\alpha;\beta}T^{\beta;\alpha}=
\Theta_{\alpha\beta}\Theta^{\alpha\beta}-
\omega_{\alpha\beta}\omega^{\alpha\beta}\ .
\ee
Note that only the vorticity term is negative.
 
    From the Ricci identity, one now has
\be
T^\alpha_{;\mu;\nu}-T^\alpha_{;\nu;\mu}=
-R^\alpha{}_{\beta\mu\nu}\, T^\beta\ ,
\ee
so
\be {d\Theta\over dt} =
(T^\alpha_{;\beta}T^\beta)_{;\alpha}-R_{\alpha\beta}\, 
T^\alpha T^\beta -
T_{\alpha;\beta}\, T^{\beta;\alpha} \ee
and hence
\be
{d\Theta\over
dt}=a^\alpha_{;\alpha}-R_{\alpha\beta}\, T^\alpha \, T^\beta-
\Theta_{\alpha\beta}\, \Theta^{\alpha\beta} +
\omega_{\alpha\beta}\, \omega^{\alpha\beta}\ .
\ee

 The assumption that the congruence is geodesic implies
$a_\alpha=0$. The assumption that it is hypersurface orthogonal
implies $T_{[\alpha;\beta}T_{\nu]}=0 \Rightarrow
\omega_{[\alpha\beta}\, T_{\nu]}=0 \Rightarrow
\omega_{\alpha\beta}=0.$ Thus,
\be
{d\Theta\over
dt}=-\Theta_{\alpha\beta}\, \Theta^{\alpha\beta}-R_{\alpha\beta}\,
T^\alpha T^\beta\ ,
\ee
and so
\be
{d\Theta\over dt}=-{1\over
d}\, \Theta^2- 2\Sigma_{\alpha\beta}\, 
\Sigma^{\alpha\beta} - R_{\alpha\beta}\, T^\alpha\, 
T^\beta\ .
\ee
Finally, we set $R_{\alpha\beta}=0$, and find ${d\Theta\over dt}<0$.

\section{Generalised Heisenberg Groups}

     As we discussed in the introduction, we are interested in the
relation between the holonomy spaces with non-abelian isometry groups
and those with nilpotent Heisenberg groups.  In this section, we
describe Heisenberg groups as contractions of semi-simple groups.

\subsection{Definition}

  We may define a generalised Heisenberg group as a (nilpotent)
central extension of an abelian group, with Lie algebra generated by
$\bfe_\a$ and $\bfq_m$, with the commutation relations
\be
[\bfe_\a, \bfe_\beta] = F_{\a\beta}^m\, \bfq_m\,,\qquad
[\bfe_a , \bfq_m]=0\,,\qquad
[\bfq_m,\bfq_n]=0\,.\label{heisalg}
\ee
We shall suppose the original abelian algebra to be $q$-dimensional,
and the centre to be $p$-dimensional.  Thus $1\le \a\le q$ and $1\le
m\le p$.  An appropriate left-invariant basis of 1-forms is $(dx^\a,
\nu^m)$, where
\be
\nu^m = dy^m -\ft12 F^m_{\a\beta}\, x^\a\, dx^\beta \,,\label{nudef}
\ee
and we have
\be
d\nu^m = -\ft12 F^m_{\a\beta}\, dx^\a\wedge dx^\beta\,.
\ee
The right-invariant Killing vectors $R_\a$ and $R_m$, which generate
left translations, are given by
\be
R_\a= \del_\a + \ft12 F_{\a\beta}\, x^\beta\, \del_m\,,\qquad
R_m = \del_m\,.
\ee
These satisfy
\be
[R_\a,R_\beta]=- F^m_{\a\beta}\, R_m\,,\quad 
[R_\a,R_m]=0\,,\quad ]R_m,R_n]=0\,.
\ee

      There is an obvious Kaluza-Klein interpretation for the central
coordinates $y^m$.  The quantities $F^m_{\a\beta}$ are just $q$
constant $U(1)$ field strengths defined over ${\Bbb E}^p$, and
(\ref{nudef}) can be viewed as the quantity $\nu^m = dy^m - A^m$,
where $A^m_\a= \ft12 F_{\a\beta}\, x^\beta$ is the Kaluza-Klein vector
potential for $F_{\a\beta}^m$.

   The Baker-Campbell-Hausdorff formula gives
\be
e^{a\cdot\bfe}\, e^{b\cdot\bfe}\, e^{-a\cdot \bfe}\, e^{-b\cdot \bfe}
= e^{\bfq_m\, F^m(a,b)}\,,\label{bch}
\ee
where we have defined $a\cdot \bfe \equiv a^\a\, \bfe_\a$ and
$F^m(a,b) \equiv F^m_{\a\beta}\, a^\a\, b^\beta$.  In practice, we
want to consider the case where the original abelian algebra is a
torus $T^q$, whose coordinates $x^\a$ therefore live on a lattice.
Because of (\ref{bch}), the coordinates $y^m$ associated to the centre
must also be identified consistently.  Suppose that $x^\a$ and
$x^\a+a^\a$ are to be identified, and that $x^\a$ and $x^\a+b^\a$ are
to be identified.  The associated group elements $e^{a\cdot\bfe}$ and
$e^{b\cdot \bfe}$ can taken to be the identity only if the group
element on the right-hand side of (\ref{bch}) is also the identity.
This means that $F^m(a,b)$ must be an integer multiple of one of the
periods of the coordinates $y^m$, for all lattice vectors $a^\a$ and
$b^\a$.  This places conditions on the $F^m_{\a\beta}$.  The case we
are mainly interested in is when the resulting group may be thought of
as a $T^p$ bundle over $T^q$.  The simplest example is when $p=1$ and
the consistency conditions reduce to the Dirac quantisation conditions
for a $U(1)$ bundle over $T^q$.  Thus in this case $F(a,b)$ is the
magnetic flux through the cycle spanned by $a^\a$ and $b^\a$.

\subsection{Heisenberg Groups as Contractions}

    Heisenberg algebras may arise as contractions of semi-simple
algebras.  The simplest example is the In\"on\"u-Wigner contraction of
$SO(3)$ to the {\it ur} Heisenberg algebra.  The former is
\be
[{\bf L}_3,{\bf L}_\pm] = \pm {\bf L}_\pm\,,\qquad
[{\bf L}_+, {\bf L}_-] = 2 {\bf L}_3\,.
\ee
Writing $\bfe_1=\lambda\, {\bf L}_1$, $\bfe_2 =\lambda\, {\bf L}_2$
and $\bfq =\lambda^2\, {\bf L}_3$, and taking the limit where the
constant $\lambda$ goes to zero, we obtain the latter:
\be
[\bfe_1,\bfe_2] = \bfq\,,\qquad [\bfe_1,\bfq]=[\bfe_2,\bfq]=[\bfq,
\bfq]=0\,.
\ee
From the Kaluza-Klein point of view, $SO(3)$ is the Dirac $T^1$ bundle
over $S^2$.  Contraction gives a magnetic field over ${\Bbb E}^2$,
which if we identify to make a torus, gives an $T^1$ bundle over
$T^2$.

   Since all the AC manifolds we are considering here have isometry
groups of the type $SO(n)$ or $SU(n)$, we shall consider the
contractions of these groups.

\subsubsection{Contractions of $SO(n+2)$}

   This example extends straightforwardly to the case of $SO(n+2)$.
It is convenient to work with the left-invariant basis of 1-forms,
which we shall denote by $L_{AB}$.  These satisfy $dL_{AB}=
L_{AC}\wedge L_{CB}$.  Splitting the index as $A=(1,2,i)$, and
defining
\be
L_{ij}= M_{ij}\,,\qquad L_{1i} = 
\sigma_i\,,\qquad L_{2i} = \td\sigma_i\,,\qquad L_{12}= 
\nu\,,\label{son2one}
\ee
then after the scalings
\be
M_{ij}\longrightarrow \lambda\, M_{ij}\,,\qquad
\sigma_i\longrightarrow \lambda\, \sigma_i\,,\qquad
\td\sigma_i\longrightarrow \lambda\, \td\sigma_i\,,\qquad
\nu\longrightarrow \lambda^2\, \nu\label{son2scale1}
\ee
we have
\bea
&&d\sigma_i = \lambda^2\, \nu\wedge \td\sigma_i + 
\lambda\, M_{ij}\wedge
\sigma_j\,,\quad
d\td\sigma_i = -\lambda^2\, \nu\wedge \sigma_i + \lambda\, 
M_{ij}\wedge\td\sigma_j\,,\quad
d\nu = -\sigma_i\wedge \td\sigma_i\,,\nn\\
&&d M_{ij} = \lambda\, M_{ik}\wedge M_{kj} - \lambda\, 
\sigma_i\wedge \sigma_j -\lambda\, \td\sigma_i\wedge \td\sigma_j\,.
\eea
Taking the limit where $\lambda$ goes to zero, we obtain the
generalised Heisenberg algebra with
\be
d\sigma_i = d\td\sigma_i =0\,,\qquad d\nu =
-\sigma_i\wedge\td\sigma_i\,,\qquad dM_{ij}=0\,.\label{t1t2n}
\ee

    The $\ft12(n+1)(n+2)$-dimensional $so(n+2)$ simple algebra has
decomposed in the $\lambda\longrightarrow0$ limit as the direct sum of
a $(2n+1)$-dimensional generalised Heisenberg algebra, spanned by
$(\sigma_i,\td\sigma_i, \nu)$, and a $\ft12n(n-1)$-dimensional
completely abelian piece, spanned by the $M_{ij}$.  It is consistent
to identify the Heisenberg group in such as way as to obtain a $T^1$
bundle over $T^{2n}$.  One has $2n$ coordinates $x^\a$ on the base,
whose differentials $dx^\a$ give $\sigma_i$ and $\td\sigma_i$.  The
real coordinates can be grouped into $n$ complex coordinates $z^i$,
with
\be
z^i= x^i +\im \, x^{i+n}\,,\qquad dz^i = \sigma_i + \im\,
\td\sigma_i\,.
\ee
The field strength $F_{\a\beta}$ is proportional to the standard
K\"ahler form on ${\Bbb C}^n$.

    A different contraction of $so(n+2)$ may be obtained by instead
applying the scalings
\be
M_{ij}\longrightarrow \lambda\, M_{ij}\,,\qquad
\sigma_i\longrightarrow \lambda^2\, \sigma_i\,,\qquad
\td\sigma_i\longrightarrow \lambda\, \td\sigma_i\,,\qquad
\nu\longrightarrow \lambda\, \nu\label{son2scale2}
\ee
After sending $\lambda$ to zero, we obtain
\be
d\sigma_i = \nu\wedge \td\sigma_i\,,\qquad d\td\sigma_i =0\,,\qquad
d\nu=0\,,\qquad dM_{ij}=0\,.\label{son2alg2}
\ee
The $\ft12(n+1)(n+2)$-dimensional $so(n+2)$ simple algebra has again
decomposed in the $\lambda\longrightarrow0$ limit as the direct sum of
a $(2n+1)$-dimensional generalised Heisenberg algebra, spanned by
$(\sigma_i,\td\sigma_i, \nu)$, and a $\ft12n(n-1)$-dimensional
completely abelian piece, spanned by the $M_{ij}$.  However now, it is
consistent to identify the Heisenberg group in such as way as to
obtain a $T^n$ bundle over $T^{n+1}$.  One has $(n+1)$ coordinates
$x^\a$ on the base manifold $T^{n+1}$ whose differentials give
$\td\sigma_i$ and $\nu$.  The field strengths $F^i_{\a\beta}$ are now
all simple; $F^i=\nu\wedge\td\sigma_i$.

\subsubsection{Contraction of $SO(4)$}

    The $so(4)$ algebra is the direct sum of two $so(3)$ algebras:
\be
d\Sigma_i = -\ft12 \ep_{ijk}\, \Sigma_j\wedge \Sigma_k\,,\qquad
d\wtd\Sigma_i = -\ft12 \ep_{ijk}\, \wtd\Sigma_j\wedge \wtd\Sigma_k\,.
\ee
Let
\bea
&&\nu_1 = \wtd\Sigma_1 -\ft12\Sigma_1\,,\qquad
\nu_2 = \wtd\Sigma_2 -\ft12\Sigma_2\,,\qquad
\nu_3 = \wtd\Sigma_3 -\ft12\Sigma_3\,,\nn\\
&&\sigma_1=\Sigma_1\,,\qquad \sigma_2=\Sigma_2\,,\qquad 
\sigma_3=\Sigma_3\,.\label{so4}
\eea
These give
\bea
&&d\nu_1 = -\nu_2\wedge\nu_3 -\ft12 \nu_2\wedge\Sigma_3 +
\ft12\nu_3\wedge \Sigma_2 + \ft14\Sigma_2\wedge\Sigma_3\,,\quad
\hbox{and cyclic on \{123\}}\,,\nn\\
&& d\sigma_1 = -\sigma_2\wedge\sigma_3\,,\qquad \hbox{and
cyclic}\,.\label{s3s3alg}
\eea

   We now implement the constant rescalings
\be
(\nu_1, \nu_2,\nu_3)\longrightarrow
\lambda^2\, (\nu_1, \nu_2,\nu_3)\,,\qquad
(\sigma_1, \sigma_2,\sigma_3)\longrightarrow
\lambda\, (\sigma_1, \sigma_2,\sigma_3)\,.\label{so4scal}
\ee
Now, we find that (\ref{s3s3alg}) becomes
\bea
&&d\nu_1 = -\lambda^2\, \nu_2\wedge\nu_3 -\ft12\lambda\,
\nu_2\wedge\sigma_3 +
\ft12\lambda\, \nu_3\wedge \sigma_2 + \ft14
\sigma_2\wedge\sigma_3\,,\qquad
\hbox{and cyclic}\,,\nn\\
&& d\sigma_1 = -\lambda\, \sigma_2\wedge\sigma_3\,,\qquad 
\hbox{and cyclic}\,.\label{s3s3alg2}
\eea
The Heisenberg limit now corresponds to sending $\lambda$ to zero,
implying
\bea
&&d\nu_1 = \ft14 \sigma_2\wedge\sigma_3\,,\qquad
d\nu_2 = \ft14 \sigma_3\wedge\sigma_1\,,\qquad
d\nu_3 = \ft14 \sigma_1\wedge\sigma_2\,,\nn\\
&&d\sigma_1=0\,,\qquad d\sigma_2=0\,,\qquad d\sigma_3=0\,.
\label{g22heis}
\eea
In this limit, we may introduce coordinates $(x_1,x_2,x_3)$ and
$(y_1,y_2,y_3)$ such that
\bea
&&\nu_1= dy_1 -\ft14 x_3\, dx_2\,,\qquad \nu_2= dy_2 - \ft14x_1\, 
dx_3\,,\qquad \nu_3= dy_3 - \ft14 x_2\, dx_1\,,\nn\\
&& \sigma_1=dx_1\,,\qquad \sigma_2 = dx_2\,,\qquad \sigma_3 = dx_3\,,
\eea
It is consistent to make identifications to give a $T^3$ bundle over
$T^3$.  The three field strengths $F^i$ are given by
\be
F^i_{jk} = -\ft12 \ep_{ijk}\,.
\ee

\subsubsection{Contraction of $SO(5)$}

    Our next example is for $SO(5)$.  Splitting the $SO(5)$ index as
$A=(\a,4)$, and then splitting $\a=(0,i)$ with $i=1,2,3$, we define
\be
L_{\a4}\equiv P_\a=\lambda^2\, 
\sigma_\a\,,\quad 
L_{0i}+\ft12 \ep_{ijk}\, L_{jk}\equiv R_i
=\lambda^4\,  \nu^i\,,\quad 
L_{0i}-\ft12 \ep_{ijk}\, L_{jk}\equiv L_i
=\lambda^3\,  J^i\,,\label{jkscal}
\ee
we obtain, after sending $\lambda$ to zero,
\be
d\nu^i=  - \sigma_0\wedge \sigma_i
-\ft12 \ep_{ijk}\, \sigma_j\wedge\sigma_k\,,\qquad d\sigma_\a=0\,,\qquad
dJ_i=0\,.
\ee
The ten-dimensional $so(5)$ algebra has thus been decomposed as a
direct sum of a seven-dimensional generalised Heisenberg algebra
spanned by $\sigma_\a$ and $\nu^i$, and a three-dimensional completely
abelian piece spanned by the $J^i$.  It is now consistent to identify
the Heisenberg group in such as way as to obtain a $T^3$ bundle over
$T^4$.  There are four coordinates $x^\a$ on the base whose
differentials give the $\sigma_\a$.  The three field strengths
$F^i_{\a\beta}$ give three self-dual 2-forms on $T^4$, which endow it
with a hyper-K\"ahler structure.  The left-invariant 1-forms can be
written as
\bea
&&\sigma_0=dx_0\,,\quad \sigma_1=dx_1\,,\quad \sigma_2=dx_2\,,\quad
\sigma_3 =dx_3\,,\\
&&
\nu_1= dy_1 -x_0\, dx_1 - x_2\, dx_3\,,\ \
\nu_2= dy_2 -x_0\, dx_2 - x_3\, dx_1\,,\ \ 
\nu_3= dy_3 -x_0\, dx_3 - x_1\, dx_2\,.\nn
\eea

    There is in fact a different contraction of $so(5)$, in which the
$J^i$ act non-trivially.  This is achieved by using the scalings
\be
L_{\a5}=\lambda^2\, \sigma_\a\,,\qquad L_{0i}+\ft12 \ep_{ijk}\, L_{jk}
=\lambda^4\,  \nu^i\,,\qquad
\qquad L_{0i}-\ft12 \ep_{ijk}\, L_{jk}
= J^i\,,\label{jkscal2}
\ee
After sending $\lambda$ to zero, we now find
\bea
&&d\nu^i= - \sigma_0\wedge \sigma_i
-\ft12 \ep_{ijk}\, \sigma_j\wedge\sigma_k\,, \nn\\
&&dJ^i=\ft12 \ep_{ijk}\,
J^j\wedge J^k\,,\nn\\
&&d\sigma_0= \ft12 J^i\wedge
\sigma_i\,,\qquad d\sigma_i = -\ft12 J^i\wedge \sigma_0 + \ft12
\ep_{ijk}\, J^j\wedge \sigma_k\,.\label{jkext2}
\eea
This algebra is the semi-direct sum of the previously-obtained
generalised Heisenberg algebra with $so(3)$, spanned by the $J^i$.

   A further contraction of $SO(5)$ is possible, leading to a
seven-dimensional Heisenberg algebra in which is the direct sum of 
a six-dimensional Heisenberg algebra and a one-dimensional summand. 
We obtain this by implementing a further singular scaling of $\nu_3$.
Equivalently, we can obtain the algebra directly as a limit of the
$so(5)$ algebra, by making the rescalings
\be
P_\a = \lambda^2\, \sigma_\a\,,\quad
R_1=\lambda^4\, \nu_1\,,\quad R_2=\lambda^4\, \nu_2\,,\quad 
R_3=\lambda^3\, \nu_3\,.\label{r1r2scal}
\ee
After sending $\lambda$ to zero, we get the contracted algebra
\be
d\nu_1=  - \sigma_0\wedge \sigma_1
-\sigma_2\wedge\sigma_3\,,\qquad
d\nu_2=  - \sigma_0\wedge \sigma_2
-\sigma_3\wedge\sigma_1\,,\qquad d\nu_3=0\,,
\qquad d\sigma_\a=0\,.\label{r1r2alg}
\ee
The left-invariant 1-forms can be written as
\bea
&&\sigma_0=dx_0\,,\quad \sigma_1=dx_1\,,\quad \sigma_2=dx_2\,,\quad
\sigma_3 =dx_3\,,\\
&&
\nu_1= dy_1 -x_0\, dx_1 - x_2\, dx_3\,,\quad
\nu_2= dy_2 -x_0\, dx_2 - x_3\, dx_1\,,\quad
\nu_3= dy_3\,.\nn
\eea

\subsubsection{Contraction of $SU(n+1)$}\label{contractappx}

    The left-invariant 1-forms $L_A{}^B$ of $SU(n+1)$ satisfy the
Maurer-Cartan relations $dL_A{}^B = L_A{}^C\wedge L_C{}^B$.  Splitting
the index as $A=(0,\a)$, we make the following definitions:
\be
L_0{}^0 =\nu\,,\qquad L_0{}^\a = 
\sigma^\a\,,\qquad L_\a{}^\beta =
 M_\a{}^\beta\,.\label{sun1}
\ee
After the scalings
\be
\nu\longrightarrow \lambda^2\, \nu\,,\qquad 
\sigma^\a\longrightarrow  \lambda\, \sigma^\a\,,\qquad
M_\a{}^\beta\longrightarrow \lambda\, M_\a{}^\beta\,,
\label{cpnscal}
\ee
and sending $\lambda$ to zero, we obtain
\be
d\nu= \sigma^\a\wedge \bar\sigma_\a\,,\qquad d\sigma^\a=0\,,\qquad
dM_\a{}^\beta=0\,.\label{cpnh}
\ee
The generalised Heisenberg algebra spanned by $\sigma^\a$,
$\bar\sigma_\a$ and $\nu$ corresponds to an $T^1$ bundle over
$T^{2n}$.  It is in fact identical to the algebra (\ref{t1t2n}) that
we obtained earlier as a contraction of $so(n+2)$.


\begin{thebibliography}{99}

\bibitem{eguhan} T.~Eguchi and A.~J.~Hanson, ``Asymptotically flat
selfdual solutions To Euclidean gravity,'' Phys.\ Lett.\ B {\bf 74},
249 (1978).

\bibitem{candossa} P.~Candelas and X.~C.~de la Ossa, ``Comments on
conifolds,'' Nucl.\ Phys.\ B {\bf 342}, 246 (1990).

\bm{brysal} R.L. Bryant and S. Salamon, ``On the construction of
some complete metrics with exceptional holonomy,'' Duke Math. J. {\bf
58}, 829 (1989).

\bm{gibpagpop} G.W. Gibbons, D.N. Page and C.N. Pope, ``Einstein
metrics on $S^3$, $\R^3$ and $\R^4$ bundles,'' Commun. Math. Phys.
{\bf 127}, 529 (1990).

\bm{hawk} S.W. Hawking, ``Gravitational instantons,'' Phys. Lett. {\bf
A60}, 81 (1977).

\bibitem{ah} M.~F.~Atiyah and N.~J.~Hitchin, ``Low-energy scattering of
nonabelian monopoles,'' Phys.\ Lett.\ A {\bf 107}, 21 (1985).

\bm{newspin7} M.~Cveti\v c, G.~W.~Gibbons, H.~L\"u and C.~N.~Pope,
``New complete non-compact Spin(7) manifolds,''
hep-th/0103155; ``New cohomogeneity one metrics with Spin(7)
holonomy,'' math.DG/0105119.

\bm{bggg} A.~Brandhuber, J.~Gomis, S.~S.~Gubser and S.~Gukov,
``Gauge theory at large $N$ and new $G_2$ holonomy metrics,''
hep-th/0106034.

\bibitem{lav1} I.~V.~Lavrinenko, H.~L\"u and C.~N.~Pope, 
``From topology to generalised dimensional reduction,''
Nucl.\ Phys.\ B {\bf 492}, 278 (1997)
[hep-th/9611134].

\bibitem{lav2} I.~V.~Lavrinenko, H.~L\"u and C.~N.~Pope,
``Fibre bundles and generalised dimensional reductions,''
Class.\ Quant.\ Grav.\  {\bf 15}, 2239 (1998)
[hep-th/9710243].

\bibitem{gibryc} G.~W.~Gibbons and P.~Rychenkova,
``Single-sided domain walls in M-theory,''
J.\ Geom.\ Phys.\  {\bf 32}, 311 (2000)
[hep-th/9811045].

\bm{cgr} M.~Cveti\v c, S.~Griffies and S.~Rey, ``Static domain walls
in N=1 supergravity,'' Nucl.\ Phys.\ B {\bf 381}, 301 (1992)
[hep-th/9201007].

\bm{cgs} M.~Cveti\v c, S.~Griffies and H.~H.~Soleng, ``Local and
global gravitational aspects of domain wall space-times,'' Phys.\
Rev.\ D {\bf 48}, 2613 (1993) [gr-qc/9306005].

\bm{cve89} M. Cveti\v c, ``Extreme domain wall - black hole
 complementarity in N=1 supergravity with a general dilaton coupling,''
Phys.\ Lett.\  B {\bf 341}, 160 (1994) [hep-th/9402089].

\bm{cs1} M. Cveti\v c and H.H. Soleng, ``Naked
singularities in dilatonic domain wall space-time,'' Phys. Rev. {\bf
D51}, 5768 (1995) [hep-th/9411170].

\bm{cs2} M. Cveti\v c and H.H. Soleng, ``Supergravity domain walls,''
Phys.\ Rep.\ {\bf 282}, 159 (1997) [hep-th/9604090].

\bm{romans2a} L.~J.~Romans, ``Massive N=2a supergravity in
ten dimensions,'' Phys.\ Lett.\ B {\bf 169}, 374 (1986).

\bm{polwit} J.~Polchinski and E.~Witten, ``Evidence for 
heterotic-Type I string duality,'' Nucl.\ Phys.\ B {\bf 460},
525 (1996) [hep-th/9510169].

\bm{berg78dual} E.~Bergshoeff, M.~de Roo, M.~B.~Green,
G.~Papadopoulos and P.~K.~Townsend, ``Duality of Type II 7-branes and
8-branes,'' Nucl.\ Phys.\ B {\bf 470}, 113 (1996) [hep-th/9601150].

\bm{gibman} G.W. Gibbons and N.S. Manton, ``Classical and quantum
dynamics of BPS monopoles,'' Nucl. Phys. {\bf B274}, 183-224 (1986).

\bm{orientifolds} N. Seiberg, ``IR dynamics on branes and space-time
geometry,'' hep-th/9606017;\\
N. Seiberg and E. Witten, ``Gauge dynamics and compactification to
three dimensions,'' hep-th/9607163;\\
A. Sen, ``Strong coupling dynamics of branes from M-theory.''
hep-th/9708002;\\
A. Sen, ``A note on enhanced gauge symmetries in M- and string
theory,''
hep-th/9707123.

\bm{mass} G.W. Gibbons and C.N. Pope, ``Positive action theorems for
ALE and ALF spaces,'' ICTP/81/82-20, available from KEK;\\
C. LeBrun, ``Counterexamples to the generalised positive action
conjecture,'' Comm. Math. Phys. {\bf 118}, 591 (1988);\\
H. Nakajima, ``Self-duality of ALE Ricci-flat 4-manifolds and positive
mass theorem,'' Adv. Studies in Pure Math. {\bf 18-I}, 385 (1990);\\
M. Dahl, ``The positive mass theorem for ALE manifolds,'' Banach
Center Publ., {\bf 41}, Part 1;\\
A. Adams, J. Polchinski and E. Silverstein, ``Don't panic! Closed
string tachyons in ALE spacetimes,'' hep-th/0108075.

\bm{clpstdomain} P.~M.~Cowdall, H.~L\"u, C.~N.~Pope, K.~S.~Stelle
and P.~K.~Townsend, ``Domain walls in massive supergravities,'' Nucl.
\ Phys.\ B {\bf 486}, 49 (1997) [hep-th/9608173].

\bm{classp}
H.~L\"u, C.~N.~Pope, T.~A.~Tran and K.~W.~Xu,
``Classification of p-branes, NUTs, waves and intersections,''
Nucl.\ Phys.\ B {\bf 511}, 98 (1998)
[hep-th/9708055].

\bm{strom} A.~Strominger and A.~Volovich, ``Holography for coset
spaces,'' JHEP 9911:013 (1999) [hep-th/9905211].

\bm{begipapo} V.A. Belinsky, G.W. Gibbons, D.N. Page and C.N. Pope,
{\sl Asymptotically Euclidean Bianchi IX Metrics In Quantum Gravity,}
Phys. Lett. {\bf B76}, 433 (1978).

\bibitem{lpsol} H.~L\"u and C.~N.~Pope,
``$p$-brane solitons in maximal supergravities,''
Nucl.\ Phys.\ B {\bf 465}, 127 (1996)
[hep-th/9512012].

\bm{cjlp1} E.~Cremmer, B.~Julia, H.~L\"u and C.~N.~Pope,
``Dualization of dualities. I,''
Nucl.\ Phys.\ B {\bf 523}, 73 (1998)
[hep-th/9710119].

\bm{hullmassive} C.~M.~Hull,
``Massive string theories from M-theory and F-theory,''
JHEP {\bf 9811}, 027 (1998)
[hep-th/9811021].

\bm{gibryc2} G.W. Gibbons and P. Rychenkova, ``Cones, tri-sasakian
structures and superconformal invariance,'' Phys. Lett. {\bf B443}, 138
(1998) [hep-th/9809158].

\bm{gibhaw} G.W. Gibbons and S.W. Hawking, {\it Gravitational
multi-instantons}, Phys.\ Lett.\ B {\bf 78}, 430 (1978).

\bm{stenzel} M.B. Stenzel, {\it Ricci-flat metrics on the
complexification of a compact rank one symmetric space}, Manuscripta
Mathematica, {\bf 80}, 151 (1993).

\bm{cglpsten} M. Cveti\v{c}, G. Gibbons, H. L\"u and C.N. Pope,
``Ricci-flat metrics, harmonic forms and brand resolutions,'' 
hep-th/0012011.

\bm{alekim}  D.V. Alekseevskii and  B.N. Kimel'fe'ld, 
''Structure of homogeneous Riemannian spaces with zero Ricci curvature
 (in Russian),''  Funkcional. Anal. i Prilo\v{z}en. 9 (1975), no. 2, 5--11. 

\bm{heber} For a review of non-compact  homogeneous
Einstein metrics, see J. Heber, ``Noncompact homogeneous 
Einstein spaces,'' Invent. Math. 133 (1998), no. 2, 279--352. 

\bm{dotti} I. Dotti-Miatello, 
``Ricci curvature of left-invariant metrics on 
solvable unimodular Lie groups,'' Math. Z. 180 (1982), no. 2, 257--263. 

\bm{guklpo} S.S. Gubser, I.R. Klebanov, A.M. Polyakov, ``Gauge theory
correlators from non-critical string theory,'' Phys.\ Lett.\ B {\bf 428}, 
105 (1998) [hep-th/9802109].

\bm{biq} O. Biquard,
``Einstein deformations of hyperbolic metrics,'' 
Essays on Einstein manifolds, 235--246, Surv. Differ. Geom.,
VI, Int. Press, Boston, MA, 1999.

\bm{stronger} R. Britto-Pacumio, A. Strominger and  A. Volovich,
``Holography for coset spaces,'' JHEP 9911 (1999) 013, [hep-th/9905211].

\bm{tayrob} M. Taylor-Robinson, ``Holography for degenerate
boundaries,'' hep-th/0001177.

\bm{gibpop} G.W. Gibbons and C.N. Pope, {\it The positive action
conjecture and asymptotically Euclidean metrics in quantum gravity},
Commun. Math. Phys. {\bf 66}, 267 (1979).

\bm{ehds} H. Pedersen, ``Eguchi-Hansen metrics with cosmological
constant,'' Class.\ Quantum Grav.\ {\bf 2}, 579 (1985).\\
H. Pedersen, ``Einstein metrics, spinning top motions and monopoles,''
Math.\ Ann.\ {\bf 274} 35 (1986).

\bm{pagpop1} D.N. Page and C.N. Pope, {\sl Inhomogeneous Einstein
metrics on complex line bundles}, Class. Quantum Grav. {\bf 4}, 213
(1987).

\bm{posasc} C.N. Pope, A. Sadrzadeh and S.R. Scuro, ``Timelike Hopf
duality and type IIA$^*$ string solutions,'' Class. Quantum
Grav. {\bf 17}, 623 (2000).

\bm{chemjomy} A. Chamblin, R. Emparan, C.V. Johnson and R.C. Myers,
``Large $N$ phases, gravitational instantons and the nuts and bolts 
of AdS holography,'' Phys.\ Rev.\ D {\bf 59}, 064010 (1999), 
[hep-th/9808177].

\bm{awacha}  A. Awad and A. Chamblin, ``A bestiary of higher
dimensional Taub-NUT-AdS spacetime,'' hep-th/0012240.

\bm{barber} M.L. Barberis, ``Homogeneous hyperhermitian metrics which
are conformally hyperk\"ahler,'' math.DG/0009035.

\bm{aspinwall} P. Aspinwall, ``M-theory versus F-theory pictures of
the heterotic string,'' hep-th/9707014.

\bm{joycebook} D. Joyce, ``Compact manifolds with special holonomy,''
(Oxford University Press, 2000).

\end{thebibliography}
\end{document}